\documentclass{aa}  

\usepackage{graphicx}
\usepackage{txfonts}
\usepackage[acronym,nomain,nonumberlist]{glossaries}
\newacronym{xifu}{X-IFU}{X-ray Integral Field Unit}
\newacronym{wfi}{WFI}{Wide Field Imager}
\newacronym{depfet}{DEPFET}{depleted field effect transistor}
\newacronym{sim}{SIM}{science instrument module}
\newacronym{spo}{SPO}{silicon pore optics}
\newacronym{fpa}{FPA}{focal plane assembly}
\newacronym[shortplural={TES}]{tes}{TES}{transition edge sensor}
\newacronym{cryoac}{CryoAC}{cryogenic anti-coincidence}
\newacronym{rmf}{RMF}{redistribution matrix file}
\newacronym{arf}{ARF}{auxiliary response file}

\newacronym{sixte}{SIXTE}{SImulation of X-ray TElescopes}
\newacronym{simput}{SIMPUT}{SIMulation inPUT}
\newacronym{fits}{FITS}{flexible image transport system}
\newacronym{hdu}{HDU}{header and data unit}
\newacronym{ra}{$\alpha$}{right ascension}
\newacronym{dec}{$\Delta$}{declination}
\newacronym{too}{ToO}{target of opportunity}
\newacronym{ew}{EW}{equivalent width}
\newacronym{fwhm}{FWHM}{full width half maximum}
\newacronym{psf}{PSF}{point square function}
\newacronym{hew}{HEW}{half energy width}
\newacronym{gcn}{GCN}{gamma-ray coordination network}
\newacronym{fov}{FOV}{field of view}
\newacronym{cie}{CIE}{collisional ionisation equilibrium}

\newacronym{cgm}{CGM}{circumgalactic medium}
\newacronym{icm}{ICM}{intra-cluster medium}
\newacronym{ism}{ISM}{interstellar medium}
\newacronym{igm}{IGM}{intergalatic medium}
\newacronym{grb}{GRB}{gamma-ray burst}
\newacronym{agn}{AGN}{active galactic nuclei}
\newacronym{whim}{WHIM}{warm-hot intergalactic medium}
\newacronym{cmb}{CMB}{cosmic microwave background}
\newacronym{fsrq}{FSRQ}{flat-spectrum radio quasars}
\newacronym{gw}{GW}{gravitational wave}
\newacronym{bh}{BH}{black hole}
\newacronym{ns}{NS}{neutron star}
\newacronym{bbn}{BBN}{Big Bang nucleosynthesis}
\newacronym{fuv}{FUV}{far-ultraviolet}
\newacronym{uv}{UV}{ultraviolet}

\newacronym{nasa}{NASA}{National Aeronautics and Space Administration}
\newacronym[longplural={European Space Agency's}, shortplural={ESA's}]{esa}{ESA}{European Space Agency}

\newacronym{athena}{\textit{Athena}}{Advanced Telescope for High ENergy Astrophysics}
\newacronym{theseus}{THESEUS}{Transient High
Energy Sky and Early Universe Surveyor}
\newacronym{svom}{SVOM}{Space Variable Objects Monitor}
\newacronym{amego}{AMEGO}{All-Sky Medium Energy Gamma-ray Observatory}

\newacronym{integral}{INTEGRAL}{International Gamma-Ray Astrophysics Laboratory}
\newacronym{gbm}{GBM}{Gamma-Ray Burst Monitor}
\newacronym{extp}{eXTP}{enhanced X-ray Timing and Polarimetry}
\newacronym{xrism}{XRISM}{X-ray Imaging and Spectroscopy Mission}
\newacronym{rgs}{RGS}{Reflection Grating Spectrometer} % On XMM
\newacronym{letgs}{LEGTS}{Low Energy Transmission Grating Spectrometer} % Chandra
\newacronym{hetgs}{HEGTS}{High Energy Transmission Grating Spectrometer} % Chandra
\newacronym{xrt}{XRT}{X-Ray Telescope}
\newacronym{cos}{COS}{Cosmic Origin Spectrograph}
\newacronym{hst}{HST}{Hubble Space Telescope}
\newacronym{epic}{EPIC}{European Photon Imaging Camera}

\newacronym{mcmc}{MCMC}{Monte Carlo Markov Chain}

\newacronym{nxb}{NXB}{non X-ray background}
\newacronym{lcdm}{$\Lambda$CDM}{$\Lambda$-Cold Dark Matter}

\usepackage{caption, subcaption, cprotect} 
\usepackage{xcolor}
\usepackage[colorlinks=true,linkcolor=blue,citecolor=blue]{hyperref}
\usepackage{cleveref}
\usepackage{fancyvrb}
\VerbatimFootnotes
\usepackage{multirow, booktabs, array}
\usepackage{float, placeins, lscape}
\usepackage{ulem, amsmath}

\newcommand{\athena}{Athena}
\newcommand{\newathena}{NewAthena}
\newcommand{\xmm}{XMM Newton}
\newcommand{\chandra}{Chandra}
\newcommand{\swift}{Swift}

\newcommand{\kev}{\text{keV}}
\newcommand{\ev}{\text{eV}}
\newcommand{\funit}{\,\text{erg}\,\text{s}\textsuperscript{-1}\,\text{cm}\textsuperscript{-2}}
\newcommand{\cm}{\text{cm}}
\newcommand{\phoind}{$\alpha_X$}
\newcommand{\nh}{$N_H$}
\newcommand{\nhsig}{$N_{H,3\sigma}$}

\begin{document} 

   \title{Future detections of the warm-hot intergalactic medium using bright power-law sources with \newathena}

   \author{J. Fisher\inst{1}
          \and
          A. Martin-Carrillo\inst{1}
          \and 
          T. Dauser\inst{2}
          \and
          J. Wilms\inst{2}
          \and
          J. Schaye\inst{3}
          \and 
          D. Barret\inst{4}
          }

   \institute{
            School of Physics, University College Dublin, Dublin 4, Ireland\\
              \email{joe.fisher@ucdconnect.ie}
        \and
            Remeis Observatory \& ECAP, Universität Erlangen-Nürnberg, Bamberg
        \and 
            Leiden Observatory, Leiden University, PO Box 9513, 2300 RA Leiden, the Netherlands
        \and 
            Université de Toulouse, CNRS, Institut de Recherche en Astrophysique et Planétologie, Toulouse
             }

   \date{Received XXX; accepted YYY}
 
  \abstract
    {Hydrodynamical cosmological simulations based on the \gls{lcdm} model predict that $\sim$40\% of the baryons in the local Universe are missing. These missing baryons are predicted to lie in low-density filamentary structures that trace connections between galaxies. This so-called \gls*{whim} is predicted to be observable in the \gls{fuv} and soft X-ray regimes, and detectable in ion species with transitions in these bands.}
    {We investigate the capability of the \gls*{xifu} on \newathena{} to detect this WHIM through \ion{O}{VII} absorption lines imprinted on bright power-law sources in soft X-rays (0.3-1.0\,keV) by examining the parameter space of different observing conditions.}
    {Through advanced simulations of the \gls*{xifu} detector, a multidimensional approach is taken to investigate the relationship between source parameters and their impacts on the significance of detection of a single absorption feature. By first studying the effect of line placement (redshift) for an observation of fixed length, followed by varying the observation length, the required counts for 3$\sigma$ and 5$\sigma$ detections are calculated for various photon indices and local Galactic absorptions. The rest-frame equivalent widths of the imprinted absorption features are then varied, and the same requirements are calculated for each line strength.}
    {We discuss the detection and recovery of the imprinted \gls*{whim} features. Another method is used to derive count requirements for detecting these features at different levels of significance, which is then compared to the previously obtained requirements. Using publicly available data from the \swift{} archives, various non-transient sources are analysed for their suitability to detect \gls*{whim} features in absorption.}
    {}


    \titlerunning{Detecting the WHIM using \newathena}
    \authorrunning{Fisher et al.}
    \maketitle
    \nolinenumbers

%%%%%%%%%%%%%%%%%%%%%%%%%%%%%%%%%%%%%%%%%%%%%%%%%%%%%%%%%%%%%%%%%%%%%%%%%%%%
% INTRODUCTION
%%%%%%%%%%%%%%%%%%%%%%%%%%%%%%%%%%%%%%%%%%%%%%%%%%%%%%%%%%%%%%%%%%%%%%%%%%%%

\glsresetall

\section{Introduction}\label{sec:introduction}

    Predictions from a combination of high-redshift \gls*{cmb} measurements \citep{Planck2020} and big bang nucleosynthesis \citep{Fields2020} demonstrate that the baryonic content accounts for $\sim$4-5\% of the energy budget in the Universe. However, in the low-redshift Universe, approximately 30-50\% of this baryonic matter is missing \citep[e.g.][]{TepperGarca2012}. Through cosmological simulations based on the $\Lambda$CDM model, it is evident that these missing baryons lie in a filamentary web that traces connections between galaxies and galactic clusters \citep{Cen2006, Cui2019, Tuominen2021}. These baryons, known as the \gls*{whim}, are shock heated through gravitational infall to high temperatures; $T$\,$\approx$\,$10^5$--$10^7$\,K, and exist in a highly tenuous state; $n_b$\,$\approx$\,$10^{-6}$--$10^{-5}\,\cm^{-3}$ \citep[e.g.][]{Cen1999, Dave2001, Nicastro2005a}. The amount of baryons residing in the \gls*{whim} increases with decreasing redshift, which is in agreement with the high-redshift observations, and explains why a high number of baryons are missing in the local Universe \citep{Cen1999}. The high temperatures of the \gls*{whim} cause its atomic constituents to be present in highly ionised states, such as \ion{H}- or \ion{He}-like ionisations, causing it to both absorb and emit in the \gls*{fuv} and soft X-ray bands. The best atomic transitions to study this matter are \ion{C}{V} in the \gls*{fuv} band, and \ion{O}{VI} and \ion{O}{VII} in the soft X-ray band, with the \ion{O}{VII}-He$\alpha$ (0.574 keV, 21.6 \AA) to be the strongest transition in this band \citep[e.g.][]{Perna1998, Cen2006, Branchini2009, Wijers2019}.
    
    Many previous attempts have been made to observe the \gls*{whim} through these absorption lines. While \citet{Fang_Mathur_Nicastro_2024} give a detailed overview, some of the milestones are described here. 
    One of the first major efforts to detect \gls*{whim} was carried along the line of sight to H1821+642 \citep{Mathur2003}. The \gls*{letgs} instrument on \chandra{} was used to detect \ion{O}{VII}-He$\alpha$, \ion{O}{VIII}-H$\alpha$ and \ion{Ne}{IX}-He$\alpha$ lines at redshifts of known \ion{O}{VI} lines. While these lines were not detected at a high significance, this method of using lines whose positions are known `a priori' provided robust methods for future work. 
    \citet{Nicastro2005a, Nicastro2005b} report on blind-line observations along the line of sight to Mrk421. Using the \gls*{rgs} on \xmm, two intervening systems at redshifts $z$\,=\,0.011 and $z$\,=\,0.026 were detected through \ion{O}{VII}-He$\alpha$ lines with different associated ions at each redshift. However, it is unclear if these absorption lines arise from filamentary \gls*{whim}, or from \gls*{cgm} of galaxies present in the line of sight \citep{Williams2010}. A follow-up of the line of sight was carried out with the \gls*{letgs} instrument on \chandra{} and no \ion{O}{VII} lines previously reported were detected \citep{Rasmussen2007}. 
    \cite{Nicastro2018} report on observations along the line of sight to 1ES1553+113 using the \gls*{rgs} instrument. They found two intervening systems; $z$\,=\,0.4339 with \ion{O}{VII}-He$\alpha$ and -He$\beta$ transitions at the same redshift as a \ion{H}{I}-Ly$\alpha$ absorption feature previously detected by the \gls*{cos} on the \gls*{hst}; and $z$\,=\,0.3551 with \ion{O}{VII}-He$\alpha$ at the same redshift as two \ion{H}{I} absorption features. Using a hybrid-ionisation model, consisting of gas in \gls*{cie} and perturbed by photoionisation, limits of the column densities, temperatures and metallicity of the absorbers were determined. However, the former system likely arises from the environment local to the host galaxy, and should be excluded from the baryon census \citep{Johnson2019, Jones2021}.
    \cite{Kovcs2019} expanded on the aforementioned work of \cite{Mathur2003} and analysed the line of sight to source H1821+643 at redshifts corresponding to known Ly$\alpha$ absorbers \citep{Tripp1998}. To probe much lower column densities than before, the spectra from four \gls*{letgs} observations were stacked at the expected locations of \ion{O}{VII} features at the previously obtained redshifts, resulting in a spectrum with an effective observation length of 8 Ms. From this, the average column density of the stacked filaments was found to be $\sim10^{15}\,\cm^{-2}$, under the assumption that all possible filaments contribute equally to the column density. It was independently verified through $10^4$ Monte Carlo simulations, which stacked 17 randomly selected redshifts in the spectrum, that this detection was statistically significant. The locations of \ion{Ne}{IX}, \ion{Ne}{X}, \ion{N}{VI}, \ion{N}{VII} and \ion{O}{VIII} were also analysed, but no significant features were detected. By adopting ionisation fractions from non-equilibrium simulations, and with the assumption that all filaments contribute equally to the detected feature, it was estimated that \ion{O}{VII} absorbers contribute $(37.5\pm10.5)\%$ of the total baryonic content.
    \cite{Ahoranta2020} used a combination of \xmm's \gls*{rgs}, and \chandra's \gls*{hetgs} and \gls*{letgs} to study absorption features along the line of sight to 3C273 at different epochs. To determine the most likely redshifts at which \gls*{whim} absorbers could be found, they examined \gls*{fuv} absorber data from \cite{Tilton2012} and determined redshifts where (1) at least two significant metal absorption lines were identified, indicating a high concentration of metals in the absorber, or (2) a broad Ly-$\alpha$ feature with $b>30\,\text{km/s}$ was found, indicating a high enough temperature ($\sim$\,$3\times10^5\,\text{K}$) to produce significant soft X-ray features. Using these criteria, they find two redshifts, $z$\,$\approx$\,0.09 and $z$\,$\approx$\,0.12, at which \gls*{whim} features were to be expected. Through simultaneous fitting of the co-added spectra of each instrument, \ion{Ne}{IX} and \ion{O}{VIII}-H$\alpha$ absorption features were found at the expected wavelengths for the $z$\,$\approx$\,0.09 system. These two ions are likely indicators of the hot \gls*{whim} phase, with temperatures $T$\,$\approx$\,$10^\text{6\,--\,6.5}$\,K. This temperature was determined using a model that assumed that the ions were in \gls*{cie} with solar abundances. Through analysis of the \gls*{ew} of the corresponding \ion{O}{VI} absorption feature using this model, it was determined that the column density of \ion{O}{VI} was underestimated, hence requiring a multi-phase structure for this to be plausible. The location at which the \ion{O}{VII}-He$\alpha$ line was predicted to fall in this system coincides with the local Galactic \ion{O}{I} line, and hence cannot be accurately studied. This highlights a key obstacle when aiming to study these features. Furthermore, no hot \gls*{whim} features were identified for the $z$\,$\approx$\,0.12 system, indicating that only a warm component exists. This lack of detection is in agreement with the line widths of the measured \ion{O}{VI} and \ion{H}{I}-Ly$\alpha$, yielding $T$\,<\,$10^{4.7}$\,K determined by \cite{Tripp2008}, and hence confirms these findings.
    
    While these older instruments have provided tentative results, it is clear that newer detectors are needed to bridge this gap. \newathena{} will provide a better opportunity to detect and study the \gls*{whim} through these absorption lines. The \athena{} mission, from which the \newathena{} mission is derived, was selected in 2014 to address and study the scientific theme of `The Hot and Energetic Universe' \citep{athena_white_paper_2013}. To do this, the reformulated \newathena{} will utilise two on-board instruments: the \gls*{xifu} \citep{Barret2023, Peille2025} and the \gls*{wfi}, the former being the instrument of use in this investigation. The \gls*{xifu} will be able to provide observations with high sensitivity to line detections, allowing detections of \gls*{whim} features in absorption \citep[e.g.][]{Wijers2020}. 
    
    Previously, for the \athena{} mission, there was a requirement of a four-hour \gls*{too} time for 50\% of observations. Hence, both \gls*{grb} afterglows and bright \gls*{agn} were selected to be used as background beacons to provide capability for absorption studies of the \gls*{whim}. Previous studies (see below) have been carried out to investigate the ability of \newathena{} to detect the missing baryons in the local Universe using \glspl*{grb} as the primary background source. However, with the current \newathena{} rescope, the \gls*{too} time goal has increased to 12 hours, making detections of \gls*{whim} through \gls*{grb} afterglows unlikely due to the rapidly decaying flux \citep{Cruise2024}. However, the findings presented in this work can still apply to the spectra of X-ray bright \glspl*{grb}, provided they produce sufficient counts during the observation. This is discussed further in \Cref{sec:grb_usage}.
    
    \citet{Walsh2020} investigated the capability to observe \gls*{whim} with older response files that assumed a larger effective area than the present setup, and utilised \glspl*{grb} as the background beacons. In each simulation, a constant photon index {\text{\phoind\,=\,2.0}} was assumed for the whole population, and each observation lasted 50 ks. The fluxes of the \glspl*{grb} were modelled using a decaying power-law model, where $F\,\propto\,t^{-\alpha}$, with $\alpha\,=\,1.2$. It was found that an excess of $10^{6}$ counts in the $0.3\,$--$\,10.0\,\kev$ range was required to yield a 75\% chance of detection of an \ion{O}{VII}-\ion{O}{VIII} pair with equivalent widths $EW_\ion{O}{VII}$\,=\,0.13\,--\,0.39\,eV and $EW_\ion{O}{VIII}$\,=\,$2/3*EW_\ion{O}{VII}$\,=\,0.09\,--\,0.26\,eV. Additionally, a limit for the local Galactic absorption of \mbox{\nh$\,<8\times10^{20}\,\cm^{-2}$} was found to be required to obtain a combined significance of $4.2\sigma$ for an \ion{O}{VII}-\ion{O}{VIII} pair, with $EW_\ion{O}{VII}$\,=\,0.28\,eV and $EW_\ion{O}{VIII}$\,=\,$2/3 * EW_\ion{O}{VII}$\,$\approx$\,0.19\,eV for a \gls*{grb} starting flux of $F_{0.3-10.0}$\,=\,$5\times10^{-11}$\funit. 
    
    These results are not accurate for \gls*{agn} due to the following; (1) The observed fluxes of \glspl{grb} have a decaying nature, as previously stated, whereas those of \gls{agn} can be modelled as approximately constant over the length of an observation of $\sim$\,50\,ks; (2) The limit was derived based on a fixed photon index $\alpha_{X,GRB}$\,=\,2.0. However, \gls*{agn} attain a range of photon indices, 1.0\,$\leq$\,$\alpha_{X,AGN}$\,$\leq$\,3.0 \citep{Joffre2022}. This range of photon indices will affect the number of photons detected in the 0.3\,--1.0\,keV band, making an updated \nh{} limit and hence the number of counts required, that are dependent on \phoind, necessary; (3) The mirror array has since been updated, and the new reduced effective area must be taken into account when accurately simulating observations. We aim to determine the source and observing parameters that are required to detect \gls{whim} absorption lines. Throughout this investigation, the background sources are not studied in detail. For simplicity, we refer to all non-transient X-ray bright power-law sources as `AGN' in this work, and they are not classified further. While we focus most of our efforts on these non-transient sources, any count or source parameter limits derived will also apply to \glspl*{grb}, provided enough counts can be observed in the short lifetime of the X-ray lightcurve decay. 
    
    The paper is broken down as follows: \Cref{sec:simulations_overview} describes the basis of the simulations, the procedures and software used, and any common details shared throughout; \Cref{sec:analysis} describes the analysis completed in this paper, and presents some minor results; \Cref{sec:results} outlines the major results from this investigation and provides a grading scheme for possible non-transient sources for use as a background X-ray beacon for detecting \gls*{whim} features, discusses measured line parameters averaged over all line redshifts, and discusses the applicability of this work for transient sources; finally, the conclusions of this work are stated in \Cref{sec:conclusions}. Throughout this work, we assume that $F$\,$\propto$\,$E^{-\alpha_X}$, where \phoind{} is the dimensionless photon index.

%%%%%%%%%%%%%%%%%%%%%%%%%%%%%%%%%%%%%%%%%%%%%%%%%%%%%%%%%%%%%%%%%%%%%%%%%%%%
% Simulations
%%%%%%%%%%%%%%%%%%%%%%%%%%%%%%%%%%%%%%%%%%%%%%%%%%%%%%%%%%%%%%%%%%%%%%%%%%%%
    
    \renewcommand{\arraystretch}{1.3}
    \begin{table}
        \begin{centering}
        \caption{Spectral parameters used to model \gls{whim} absorption features in continuum of bright X-ray sources, unless otherwise stated.}    
        \label{tab:sim_params}
        \begin{tabular}{llc}\hline
            Component & Parameter & Reference \\ \hline\hline
            \textbf{Local Absorption} & $10^{20}$\,--\,$10^{22}$\,cm$^{-2}$ &  \\ \hline
            \textbf{X-ray power-law} &  &  \\
            Flux (0.3-10.0 keV) & $5\times10^{-11}\funit$ &  \\
            Photon Index (\phoind) & 1.0\,--\,3.0 & 1 \\ \hline
            \textbf{\ion{O}{VII} Absorption Line} &  &  \\
            Rest Energy ($E$) & 0.574 keV &  \\
            Redshift ($z$) & 0.0\,--\,0.5 & 2 \\
            Rest EW  & 0.28\,eV & 3 \\
            Rest Line Width ($\sigma$) & 0.1\,eV &  4\\\hline
        \end{tabular}
        \end{centering}
    \tablebib{(1) \cite{Joffre2022}; (2) \cite{Walsh2020}; (3) \cite{Nicastro2018}; (4) \cite{Brand2016}}
    \end{table}

\section{Simulating WHIM absorption spectra}\label{sec:simulations_overview}

    \gls*{sixte}, the official Monte Carlo simulation toolkit for the \gls*{xifu}, was used to realistically simulate the observations in this investigation \citep{Dauser2019}. This work adopted the current baseline response files\footnote{ARF: athena\_xifu\_13\_rows\_no\_filter.arf\\XML: xifu\_nofilt\_defoc.xml\\PSF: athena\_psf\_defoc\_20240326.fits\\NXB: athena\_xifu\_nxb\_sixte.pha} as described by \citet{Peille2025}, that assumes an effective area of $\sim$\,$0.62\,\text{m}^2$ at 1.0 keV with no filters. The geometry of the instruments, as well as the capability of the mirror to provide a 35\,mm defocusing configuration, were modelled throughout the simulations to allow for maximum photon collection and to prevent saturation of pixels. 

    The required inputs for the pipeline are the XML file that describes the \gls*{xifu} calorimeter geometry and the grading scheme for the generated photons, and the \gls*{simput} file \citep{Schmid2013} that describes the spectrum, light curve and position of the observed source. \verb|XSPEC v12.14.1|, interfaced using \verb|pyxspec v2.1.4|, was implemented to create models of the source spectra. Other inputs for the pipeline included the telescope pointing, the exposure time of the observation, and whether crosstalk between pixels should be simulated. In all cases, the pointing was set to the source position, such that it was centred and defocused on the detector. Crosstalk was simulated between pixels throughout, and a physically accurate \gls*{nxb} was included to reflect the baseline configuration of the detector. Due to the high effective area of \newathena{}, a cosmic X-ray background was not considered here as its effect is negligible in comparison to the count rates from sources utilised here (see \cite{Lotti2021} for predicted contributions). Only photons with the highest energy resolution ($\Delta E\,\leq4\,$eV) were used for spectrum generation, and all spectra remained unbinned for analysis.

    \begin{figure*}[ht]
        \centering
        \includegraphics[width=\linewidth]{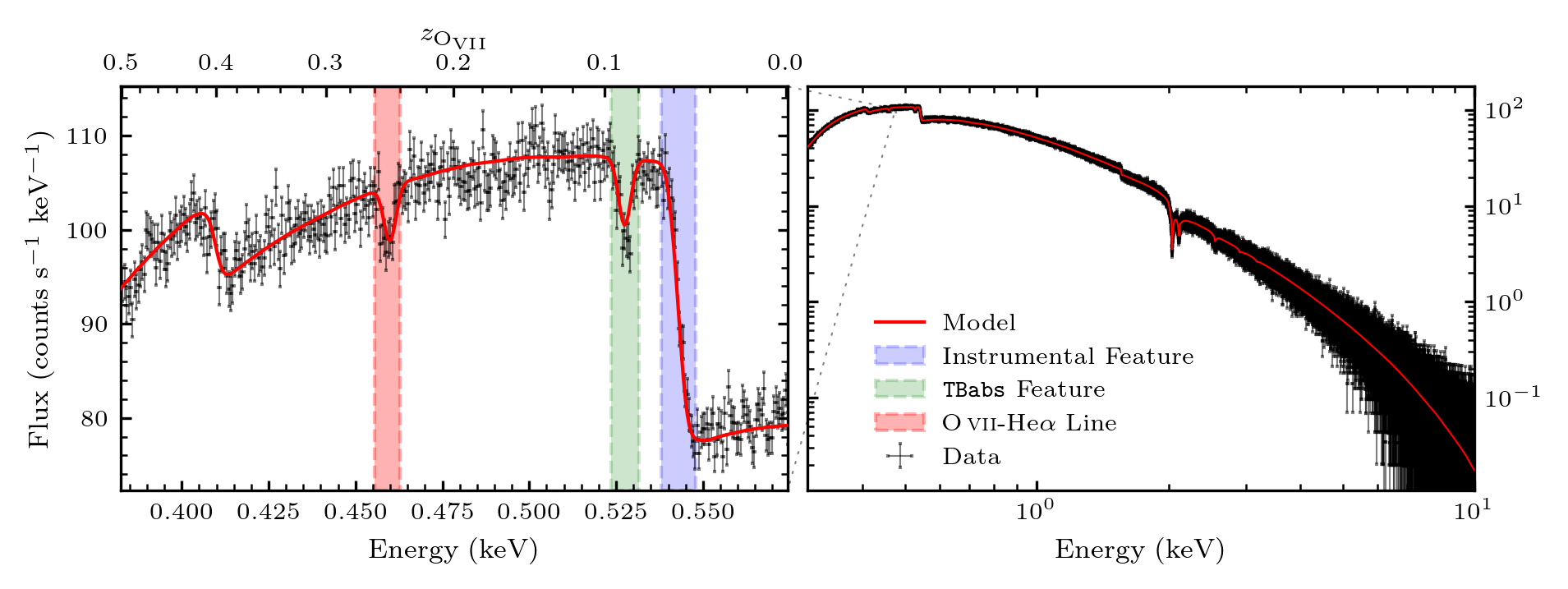}
        \cprotect\caption{50\,ks simulated \gls*{xifu} observation with photon index \phoind\,=\,2.0, local Galactic \nh\,=\,$1\times10^{20} \,\cm^{-2}$, and unabsorbed source flux $F_{0.3-10.0\,\kev}$\,=\,$5\times10^{-11}\funit$. \textit{Right:} Simulated spectrum over the full 0.3\,--\,10.0\,\kev energy range. \textit{Left:} Spectrum in the 0.3\,--\,1.0\,$\kev$ energy range. An imprinted \ion{O}{VII}-He$\alpha$ absorption feature with rest-frame \gls*{ew} of 0.28\,eV can be seen at $z$\,=\,0.25 (red band). The instrumental feature and local absorption feature contained in the \verb|TBabs| model are highlighted in blue and green, respectively.}
        \label{fig:example_whim_spectrum}
    \end{figure*}
    
    All spectral parameters were set to the values seen in \Cref{tab:sim_params} unless otherwise stated. The observed continuum spectra of the sources, absent of any \gls*{whim} absorption features, consist of two components: cold local Galactic \gls*{ism} absorption in the Milky Way, represented by \verb|TBabs| \citep{Wilms2000}, and a simple power-law representing the emission of the source, \verb|pegpwrlw|. To model a \gls{whim} absorption line, the \verb|zgauss| model with a negative normalisation parameter was added to the continuum, which represented an \ion{O}{VII}-He$\alpha$ absorption feature with rest-frame energy $0.574\,\kev$ ($21.6\,\AA$). Throughout this investigation, only single-line absorption spectra were studied, with the rest-frame \glspl{ew} of the lines fixed to 0.28\,eV ($\sim$\,10.5\,m\AA), representing a mid-strength line; \cite{Wijers2019} predict a possible maximum \gls*{ew} of 0.42\,eV, see also \cite{Walsh2020}. The rest-frame line width was set to $\sigma$\,=\,0.1\,eV throughout \citep{Brand2016}. Holding these parameters fixed for the initial investigation allowed a study of the effects of line placement in the spectrum. It also provided an opportunity to investigate the effect of power-law spectral parameters, which is important when aiming to detect these absorption features in the spectra of bright transient background sources, such as \glspl{grb}, as they may not be quantified. The line \gls*{ew} was varied in the later stages of the work. The source fluxes were modelled as constant over the length of each simulation, and were set to $F_{0.3-10.0\,\text{keV}}$\,=\,$5\times10^{-11}\funit$ throughout, unless otherwise stated. Keeping the flux constant allowed the observation length to serve as the primary variable governing the detected counts, rather than a combination of flux and observation length. All sources were simulated and fit over the full 0.3\,--\,10.0\,keV energy range using Cash statistics \cite{Cash1979}.
    \Cref{fig:example_whim_spectrum} displays an example \gls*{whim} absorption spectrum simulated using these observing and source parameters, with a line redshift of $z$\,=\,0.25. The \ion{O}{I}-K$\alpha$ transition contained in \verb|TBabs| can be seen at $E\sim527.5\,\ev$ ($\sim$\,23.5\,\AA). Here, we do not consider the warm or hot phases of the local Galactic \gls*{ism}. This is because the non-cold \gls*{igm} is very line-of-sight dependent, and is not yet well-quantified. This makes it harder to accurately represent without adding two more dimensions for analysis for the column densities of the warm and hot phases. Moreover, while the inclusion of such absorbers would affect the early results of this study, the final results are independent of these quantities, discussed further in \Cref{sec:analysis}.

    \begin{figure}[t]
        \centering
        \includegraphics[width=\linewidth]{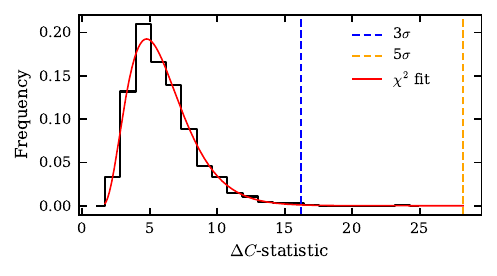}
        \caption{Example of an obtained $\Delta\mathcal{C}$-stat distribution used to calculate line significances. The 3$\sigma$ and 5$\sigma$ thresholds are shown.}
        \label{fig:deltaC_distribution}
    \end{figure}

%%%%%%%%%%%%%%%%%%%%%%%%%%%%%%%%%%%%%%%%%%%%%%%%%%%%%%%%%%%%%%%%%%%%%%%%%%%%
% Analysis
%%%%%%%%%%%%%%%%%%%%%%%%%%%%%%%%%%%%%%%%%%%%%%%%%%%%%%%%%%%%%%%%%%%%%%%%%%%%

\section{Data analysis}\label{sec:analysis}

    \begin{figure*}[h]
        \sidecaption
        \includegraphics[width=0.65\linewidth]{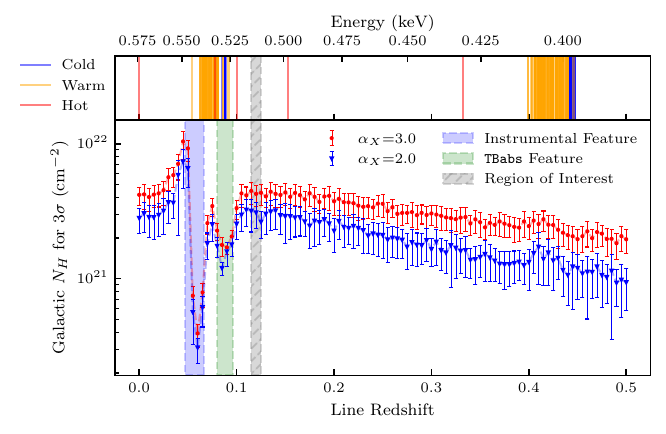}
        \caption{Redshifts of \ion{O}{VII}-He$\alpha$ absorption features against the \nh{} that corresponds to a 3$\sigma$ detection. Regions affected by instrumental features and local absorption in the Milky Way (contained in TBabs model) are highlighted by the blue and green regions, respectively, as seen in \Cref{fig:example_whim_spectrum}. Data for \phoind\,=\,1.0 have been excluded due to lack of detection. The upper portion of the figure displays the energies of known cold, warm and hot \gls*{igm} ions, of which only \ion{O}{I} is included in this work \citep{Gatuzz2023, Mendoza2021}. \vspace{2.0cm}}
        \label{fig:zregion_vs_galnh_50ks}
    \end{figure*}

    \subsection{Investigation of line placement}\label{sec:first_sims}

    The initial investigation of this work involved determining the optimal line placement for detecting \gls{whim} absorption lines for a fixed observation length of 50\,ks and fixed unabsorbed source flux of $F_{0.3-10.0\,\text{keV}}$\,=\,$5\times10^{-11}\funit$. Holding these fixed reduced the number of simulations required to investigate the relation between observing parameters and the significance at which \gls{whim} absorption lines are detected. Only the cold Galactic absorption (hereafter \nh), photon index (hereafter \phoind), and redshift of the line ($z$) were free to vary. These parameters served as the main dependent variables when detecting \gls{whim} features in absorption in the spectra of bright power-law sources. For non-transient sources such as \gls{agn}, these parameters are well-known. However, for transient sources such as \glspl{grb}, these parameters will be unknown. Hence, it is of utmost importance to determine under what observing conditions a \gls{whim} feature can be detected. With the unabsorbed flux of the source fixed, the continuum is defined solely by \phoind{} and \nh. \phoind{} values of 1.0, 2.0 and 3.0 were used, and \nh{} was varied between $1\times10^{20}\,\cm^{-2}$ and $1\times10^{22}\,\cm^{-2}$ in nine equal log-spaced steps.
    The redshift of the line was varied between 0.0 and 0.5 in 101 linear steps. For each combination of \phoind, \nh, and $z$, 25 spectra were simulated using \gls{sixte} and analysed in \verb|XSPEC|. Each spectrum was fit with the same model with which it was initialised (\verb|TBabs*(pegpwrlw + zgauss)|). It was assumed that the positions of the lines are known `a priori'\footnote{i.e. the positions of the lines are already known, through detections of other features along the line of sight to the source, such as Ly$\alpha$ or \ion{O}{VI} absorption features.} \citep[e.g.][]{Spence2025}, thus we did not carry out any blind-line fitting procedures in this work. It was assumed, as is in a real observation, that the \nh{} along the line of sight to the source is known. The continuum-only model was thus initialised with a fixed \nh{} corresponding to the known value, and with \phoind{} and flux of the power-law source free to vary, starting from their true values. While this is not accurate for a true observation, any effects from a poorly fit continuum are removed in this work. The $\mathcal{C}$-stat corresponding to the best-fitting model of the continuum was recorded. A Gaussian absorption line model was added with a rest energy corresponding to that of the \ion{O}{VII}-He$\alpha$, and its redshift was frozen at its known value. This leaves the negative line normalisation and sigma to vary. To obtain the best fit, \verb|steppar| was used, and the test statistic corresponding to the best fitting line model, $\mathcal{C}_{\text{line}}$ was recorded.
    
    To assess the measured significance of an absorption feature, Monte Carlo simulations were carried out to eliminate the chance that the measured lines are affected by statistical fluctuations \citep{Protassov2002, Walsh2020}. For each simulated observation, 1000 spectra were simulated using the best-fitting models with no absorption features present, which was assumed as the null hypothesis model. To quantify the statistical fluctuations present in the spectrum, a single Gaussian absorption feature was implemented as before. The energy was varied between 0.3\,--\,1.0\,keV, and a fit was performed. The continuum $\mathcal{C}$-stat and the minimum $\mathcal{C}$-stat from the fluctuation search were measured. A distribution of the $\Delta\mathcal{C}$ values was then obtained;
    \begin{equation}\label{eqn:deltaC}
        \Delta C = C_{\text{cont}} - C_\text{cont+line}.
    \end{equation}
    In the limit of many counts per bin (which can be assumed given the large effective area of \newathena), this quantity acts asymptotically as a $\chi^2$ distribution \citep{Cash1979}. Using this distribution, the statistical significance of the absorption features was calculated. \Cref{fig:deltaC_distribution} displays an example distribution used to calculate line significance.
    
    For each \phoind, \nh, and line $z$, the mean significances of the line detections and their associated uncertainties were calculated assuming a Gaussian distribution across the 25 simulated spectra. To compare the effects of line placements across various \phoind{} and \nh{} values, we calculated the \nh{} that corresponds to a 3$\sigma$ detection (hereafter \nhsig) for each line $z$ and \phoind. This was done using Monte Carlo fitting methods to determine the relationship between \nh{} and measured line significance, from which \nhsig{} was calculated. This quantity aids in the comparison between different \phoind{} and $z$ values; a higher \nhsig{} implies a larger measured line significance across all values of \nh. Further details of calculations and discussions are presented in \Cref{sec:galnh_appendix}. \Cref{fig:zregion_vs_galnh_50ks} displays the calculated \nhsig{} values across the redshift window considered. Absorption features imprinted on power-laws with \phoind\,=\,1.0 were not detected at a high enough level of significance for redshifts outside of the instrumental and cold local absorption features (in blue and green, respectively), and hence these data have been excluded. The upper portion of the figure displays the positions of known lines which correspond to the cold (\ion{N}{I}, \ion{O}{I}, \ion{Ne}{I}), warm (\ion{N}{II}, \ion{N}{III}, \ion{O}{II}, \ion{O}{III}, \ion{Ne}{II}, \ion{Ne}{III}) and hot (\ion{N}{IV}, \ion{O}{VII}, \ion{Ne}{IX}) local \gls*{igm} \citep{Gatuzz2023, Mendoza2021}. The expected column densities of the warm and hot components along the lines of sight to extragalactic sources are much lower than the measured cold column densities \citep{Gatuzz2017}, and hence have a much lower effect on the observed spectra.
    We identify the redshift region $z$\,=\,0.110\,--\,0.125, highlighted in grey, as a region applicable for further analysis of the effect of the background power-law parameters on the measured line significance. This is due to two things: (1) This region is not affected by the instrumental feature and the local \ion{O}{I} absorption feature as modelled in this analysis; (2) It is the region with the highest values of \nhsig{} with $z$\,>\,0.1; and (3) This region is not affected by the warm and hot \gls*{igm} components absent from this analysis (see the upper portion of \Cref{fig:zregion_vs_galnh_50ks}), thus we need no longer consider their poorly-constrained column densities.
    
    Near the redshifts of the included cold \gls{igm} feature and the instrumental feature, \nhsig{} varies drastically, implying poor line detection in these regions. This is the result of incorrect significance measurements due to these features (see \Cref{sec:galnh_appendix}). Moreover, by including features arising from the warm and hot \gls{igm}, these significance measurements will worsen again. Hence, we would fail to detect \ion{O}{VII}-He$\alpha$ lines in absorption with redshifts $z$\,<\,0.1 with reliable significance measurements.

    \begin{figure}[t]
        \centering
        \includegraphics[width=\linewidth]{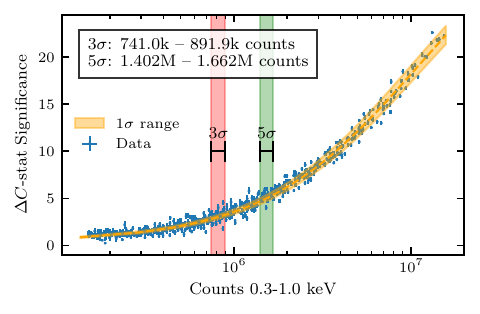}
        \caption{Relation between counts (0.3\,--\,1.0\,keV) and line significance for lines with $z$\,=\,0.110\,--\,0.125. The counts required for 3$\sigma$ and 5$\sigma$ detections are highlighted in red and green, respectively.}
        \label{fig:counts_for_sigs}
    \end{figure}

    \subsection{Derived count limits}\label{sec:requiredCounts}

    The most direct proxy for determining the expected line significance is the counts detected in the energy band of interest. To accurately determine the relation between the significance of detection of a \gls*{whim} absorption feature and the counts detected for lines of a rest-frame \gls*{ew} of 0.28\,eV, the simulated observation length was varied between 6.25\,ks and 150.0\,ks. The same procedure for simulation as outlined in \Cref{sec:simulations_overview} was used, with the addition of more \phoind{} values; between 1.0 and 3.0 with a step size of 0.25. Only lines in the range $z$\,=\,0.110\,--\,0.125 were simulated, which removed the effects from the instrumental and \verb|TBabs| absorption features, as well as the effects from the warm and hot \gls*{igm} not included here, on the measured line significance. It also provided upper limit results due to the high \nhsig{} in the redshift range. For each line simulated, the line significance was calculated, and the counts detected in the 0.3\,--\,1.0\,keV range were recorded. This energy range was selected as it includes the transition energies of ions that will be used to detect and study the \gls*{whim} (e.g. \ion{O}{VII}, \ion{O}{VII}, \ion{Ne}{IX} and \ion{Fe}{XVII}), while removing energies (>1\,keV) in which detected counts that will not directly contribute to the detected significance of these ions. \Cref{fig:counts_for_sigs} displays the relation between counts and line significance of all simulations carried out. Each point represents the mean observed counts and mean line significance, calculated using \Cref{eqn:deltaC}, and their associated uncertainties from across the 25 simulated spectra. The expected single line significance within $1\sigma$ uncertainties, given that $n$ counts have been detected, was determined through
    \begin{equation}\label{eqn:sigs_counts_relation}
        \sigma(n) = \frac{an}{b+n}+c.
    \end{equation}
    While this functional form is not physically motivated, it provided excellent fits of the data. The best-fit parameters were obtained using Monte Carlo fitting methods as before. Using this, it was found that $\sim$\,800\,k counts are required to detect a single line at a $3\sigma$ level, and $\sim$\,1.5\,M counts at a $5\sigma$ level for a line of rest-frame \gls*{ew} of 0.28\,eV.

    Due to the high dependency between measured line significance and line \gls*{ew}, it is of crucial importance that count limits are derived for varied line strengths in the same redshift region. This process was thus repeated for lines with rest-frame \glspl*{ew} of {0.14\,--\,0.42\,eV}. Observations of the varied absorption features were simulated for exposure times of 25, 50, 100 and 150\,ks, with \phoind{} values between 1.0\,--\,3.0 with a step size of 0.5 and the same \nh{} values as before. \Cref{eqn:sigs_counts_relation} was used to calculate the count requirements for detection at the 3$\sigma$ and 5$\sigma$ level as before. The results are displayed and discussed in \Cref{sec:count_limits_first_results}.

    \subsection{Analysing real sources}\label{sec:real_sources}
    
    Many sources, both of transient and non-transient nature, are predicted to provide adequate brightness for detecting \gls{whim} features in absorption. Here we focus our efforts on studying non-transient sources, simply referred to here as \gls{agn}, and provide a grading scheme for their suitability. To do this, their fluxes in the energy band of interest and their photon indices must be examined for different observing lengths given the known \nh{} along the line of sight to the source to ensure that they can detect enough counts ($\sim$\,800\,k for a $3\sigma$ detection, and $\sim$\,1.5\,M for $5\sigma$) in a reasonable observation length. 

    The \gls{agn} that were investigated in this work were selected due to their inclusion in previous works; \citet{Fang2005} analysed six different sources for \gls{whim} absorption features in the X-ray regime, but did not find any lines present across observations from various instruments; \citet{Danforth2010} studied thermally broadened Ly$\alpha$ absorbers in the lines of sight to seven background sources using \gls{hst}; \citet{Prochaska2011} investigated the connection between \gls*{igm} and galaxies by investigating the fields of 20 UV-bright quasars; \citet{Tilton2012} studied absorption lines from ions in the UV regime, including \ion{O}{VI} and \ion{C}{VI} to account for missing baryonic matter in the local Universe; \citet{Savage2014}, later updated by \citet{Stocke2014}, investigated \ion{O}{VI} and \ion{H}{I} absorption features along the line of sight to 14 quasars; \citet{Bregman2015} compiled a list of background sources for detecting \gls{whim} features in absorption, and provided a method to quantify the merit of each source; \citet{Dalton2021} studied the properties of intervening \gls{igm} along the lines of sight to 40 randomly selected blazars; \citet{Jones2021} developed a method for constraining the redshifts of blazars, particularly those which will provide an opportunity to detect \gls{whim} absorption features; \citet{Spence2025} performed a systematic search for X-ray absorption features in the spectra of sources in which UV absorption features have previously been detected. \gls{agn} from these previous works with $z$\,>\,0.1 were selected for analysis in this work. The full selection of sources can be seen in \Cref{sec:full_sources_appendix}.
    
    Archival data of the selected \gls{agn} were analysed from the \swift database \citep{Swift_Evans2009}. For each source, the time-averaged spectra across all PC-mode observations were created. Using these, the flux between 0.3\,--\,10.0\,keV and \phoind{} were extracted by fitting the spectrum with an absorbed power-law model, allowing the \phoind{} and flux to vary until a fit was reached. The \verb|TBabs| \nh{} value was frozen at its known value along the line of sight to the source. Using \verb|steppar|, the two-dimensional relation between the \phoind{} and flux was found, with the \nh{} frozen at its best-fit value. The limiting ellipse that corresponds to $\Delta C \sim \Delta\chi^2 = 9.21$ was extracted in each case, providing the source state at a 99\% confidence level, averaged across all observations. Along this ellipse boundary, the observed \swift{} count rates were calculated, and \phoind{} and flux values corresponding to the maximum and minimum count rates were extracted. Using these values, the expected \gls{xifu} count rate was calculated by simulating ten 25\,ks observations with the \phoind{} and flux set to those corresponding to the best fit, the minimum and the maximum expected count rates.

    To allow for comparison between the \gls{agn} and to determine which are better suited to study the \gls{whim} features in absorption, we adapted the methods of \cite{Bregman2015}, where the merit of a source can be calculated using
    \begin{equation}\label{eqn:merit}
        M = zC(N_{H}, \alpha_X, F_3X),
    \end{equation}
    where $C(N_H, \alpha_X, F_X)$ is the observed \gls{xifu} counts in the $0.3-1.0\,\kev$ energy band obtained from the ellipse-fitting method. We assumed that $dN(\ion{O}{VII})/dz$ is constant at low redshifts \citep[figure 3]{Wijers2019}. $z$ was taken as the minimum between $z_{AGN}$ and unity, above which \ion{O}{VII}-He$\alpha$ features cannot be detected, and hence the weighting for filaments above this redshift was removed. This modification reduced the energy range used to calculate the merit, reflecting more accurately the expected energies of \gls{whim} absorption features and incorporated the spectral shape as well as the flux of the source. This metric, while unphysical, favours sources that can provide high counts in a shorter observation time, as well as those which probe longer lines of sight and thus have a higher chance of probing filaments. The full results are discussed in \Cref{sec:source_grading_results}.

    This investigation did not account for different states of the \gls{agn} (i.e. flaring), and only used their time-averaged states as observed by \swift. Flaring causes an increase in observed flux, but is often correlated to a decrease in \phoind{} (i.e. an increase in hardening). This so-called harder-when-brighter behaviour has been seen in sources included in \Cref{tab:full_sources_table}, for example PKS2155-304 \citep{Bhagwan2016}; and other sources not studied here; Mrk501 \citep{Krawczynski2000}; Mrk421 \citep{Abeysekara2016, Abe2025}. \citep[e.g.][]{Zhang2002} However, other sources have displayed a weakly correlated softer-when-brighter behaviour during flaring; PKS0716+714 and 3C66A \citep{Wierzcholska2016}; and others \citep[e.g.][]{Majumdar2025}. In the case of 3C273, the correlation between \phoind{} and flux has been observed to be positive, negative and absent over the course of several years \citep{Soldi2008}. For these reasons, flaring states are not considered, as their complex natures are beyond the scope of this work.
        
%%%%%%%%%%%%%%%%%%%%%%%%%%%%%%%%%%%%%%%%%%%%%%%%%%%%%%%%%%%%%%%%%%%%%%%%%%%%
% Results & Discussion
%%%%%%%%%%%%%%%%%%%%%%%%%%%%%%%%%%%%%%%%%%%%%%%%%%%%%%%%%%%%%%%%%%%%%%%%%%%%

\section{Results and discussion}\label{sec:results}

    \subsection{Line placement effects}

    Mid-strength \gls*{whim} absorption features, in the form of \ion{O}{VII} lines of rest-frame \gls*{ew} 0.28\,eV, have been accurately represented in the continuum of bright X-ray power-law sources using the \newathena{} \gls*{xifu} instrument with the 35\,mm mirror defocusing capability, as detailed in \Cref{sec:simulations_overview}. To test the effects of line redshift on the measured line significance, advanced simulations using \gls*{sixte} toolkit were carried out. By assuming a fixed observation length of 50\,ks and unabsorbed source flux of $F_{0.3-10.0}$\,=\,$5\times10^{-11}$\funit, X-ray background sources were simulated with photon indices 1.0--3.0 and local Galactic absorption $10^{20}$\,--\,$10^{22}$\,cm$^{-2}$ modelled using \verb|TBabs|. The line redshifts were varied in equal steps of size $dz$\,=\,0.005 for $z$\,=\,0\,--\,0.5. For each redshift, the \nh{} value that corresponds to a $3\sigma$ detection, \nhsig, was calculated, as seen in \Cref{fig:zregion_vs_galnh_50ks}. No \nh{} values yielded a significant detection for \phoind\,=\,1.0. It can be seen that at redshifts near the instrumental feature and local \verb|TBabs| Galactic absorption, the measured line significances, and hence \nhsig{} was reduced. Due to warm and hot \gls*{igm} lines not modelled in this work (see \Cref{fig:zregion_vs_galnh_50ks}), a detection of \ion{O}{VII}-He$\alpha$ absorption features arising from the \gls*{whim} with $z$\,<\,0.1 is unlikely. At redshifts $z$\,>\,0.1, an approximately linear decrease of \nhsig{} with increasing $z$ was observed. 

    \subsection{Minimum required counts}\label{sec:count_limits_first_results}

    Following the initial simulations of varied redshift regions, a relation between counts and measured line significance was found to determine the minimum counts required for a significant detection. To do this, the redshifts of lines were set to values between $z$\,=\,0.110--0.125, where \nhsig, as seen in \Cref{fig:zregion_vs_galnh_50ks}, remains approximately constant and is not affected by local absorption or instrumental features. \Cref{fig:counts_for_sigs} displays the relation between observed counts in the 0.3--10.0\,keV range and the measured line significance, which is well described by \Cref{eqn:sigs_counts_relation}. Using this relation and its associated 1$\sigma$ uncertainty, for a \gls{whim} absorption feature with rest-frame \gls*{ew} of 0.28\,eV and redshifts $z$\,=\,0.110\,--\,0.125, it was found that $\sim1.5\times10^6$ counts in the 0.3--1.0\,keV range are required for a 5$\sigma$ detection, and $\sim8\times10^{5}$ counts are required for a 3$\sigma$ detection. While the range in which the counts detected could have been decreased to better match the possible locations of where the \ion{O}{VII} absorption features falls (that is, $\sim$0.38\,--\,0.58\,keV corresponding to $z$\,=\,0.0\,--\,0.5), the larger band was chosen to account for other transitions not included in this work, such as \ion{O}{VIII}, \ion{Ne}{IX} and \ion{Fe}{XVII}. 

\begin{figure}[t]
        \centering
        \includegraphics[width=\linewidth]{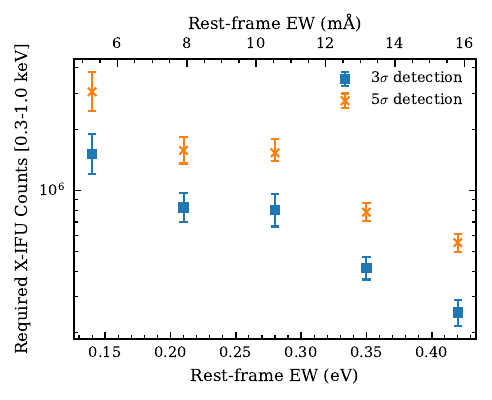}
        \caption{Required \gls*{xifu} counts (0.3\,--\,1.0\,keV) to provide a 3$\sigma$ and 5$\sigma$ detection of lines with different rest-frame \glspl{ew} and $z$\,=\,0.110\,--\,0.125.}
        \label{fig:varied_ew_counts}
    \end{figure}

    This process was repeated for lines of rest-frame \glspl*{ew}\,=\,0.14\,--\,0.42\,eV to quantify the count requirements for a detection at the 3$\sigma$ and 5$\sigma$ level. \Cref{fig:varied_ew_counts} displays the results from this analysis. The variation related to the count requirement for a 0.28 eV line arises from the fact that more simulations were carried out compared to the other \glspl{ew}. As expected, fewer counts are required for more prominent lines, whereas more counts are required for weaker lines, provided that they are detected in the $z$\,=\,0.100\,--\,0.125 region. Thus, to probe weaker lines, \newathena{} should target: (1) bright non-transient sources that can provide enough counts in a feasible exposure time, or dimmer sources provided a longer exposure is possible; or (2) transient sources that can provide a high flux over a short period, given that the \gls{too} response time is sufficient.
    
    \subsection{Source grading}\label{sec:source_grading_results}

    \begin{figure*}[h]
        \centering
        \includegraphics[width=\linewidth]{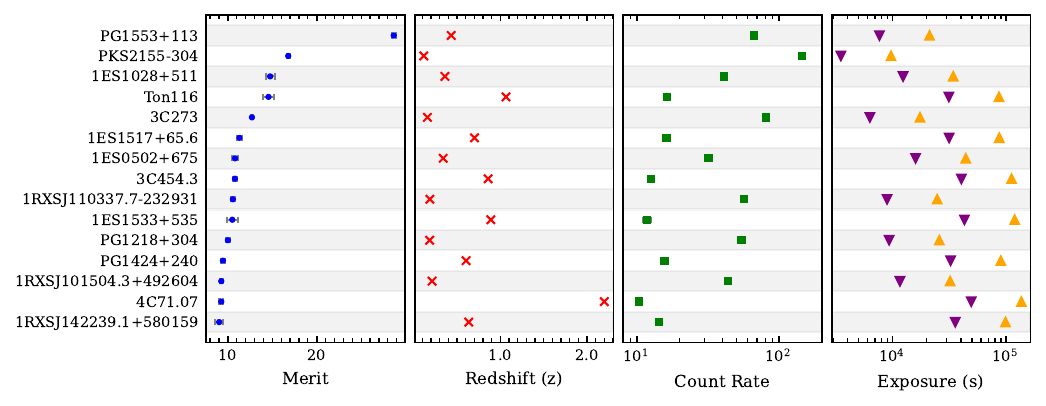}
        \caption{Sources investigated in this work with the 15 highest merit values, displayed with source redshift and measured \newathena{} count rate (0.3\,--\,1.0\,keV). \Cref{fig:appendix_merit_sources} displays the results for all sources. The rightmost portion displays the required observation length to detect a feature at 3$\sigma$ (purple, downward) and 5$\sigma$ (orange, upward) levels using the count limits from \Cref{fig:counts_for_sigs}.  All data can be seen in \Cref{tab:full_sources_table}.}
        \label{fig:merit_sources}
    \end{figure*}

    Real non-transient X-ray power-law sources, referred to in this work simply as \gls{agn}, that have been used in previous work (see \Cref{sec:real_sources}), were analysed to determine their suitability to use as background beacons for detecting and studying \gls*{whim} lines in absorption. Only \gls{agn} with $z$\,>\,0.1 were considered, and it was assumed that $dN/dz$ remained constant for low redshifts. \Cref{fig:merit_sources} displays a sample of sources ordered by their calculated merits, along with the source redshift and expected \gls{xifu} count rates in the 0.3\,--\,1.0\,keV range. We also include the required observation length to detect a feature with a rest-frame \gls{ew} of 0.28\,eV, as calculated in \Cref{sec:count_limits_first_results}. The full results can be seen in \Cref{sec:full_sources_appendix}.

    \begin{figure*}[t]
        \centering
        \includegraphics[width=\linewidth]{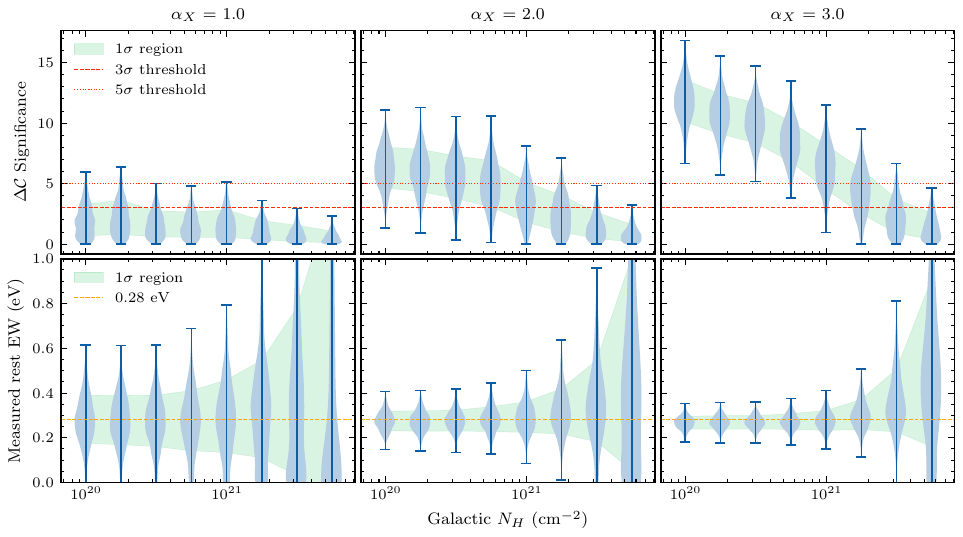}
        \caption{Derived line properties averaged across all lines with $z$\,>\,0.1. The 1$\sigma$ uncertainty is highlighted in green. \textit{Top:} Measured line significances for different \nh values. The 3$\sigma$ (dashed line) and 5$\sigma$ (dotted line) are displayed. \textit{Bottom:} Recovered rest-frame \gls{ew} of lines with input \gls{ew} of 0.28 eV (dashed line).}
        \label{fig:line_measures}
    \end{figure*}

    \begin{figure*}[ht]
        \centering
        \includegraphics[width=\linewidth]{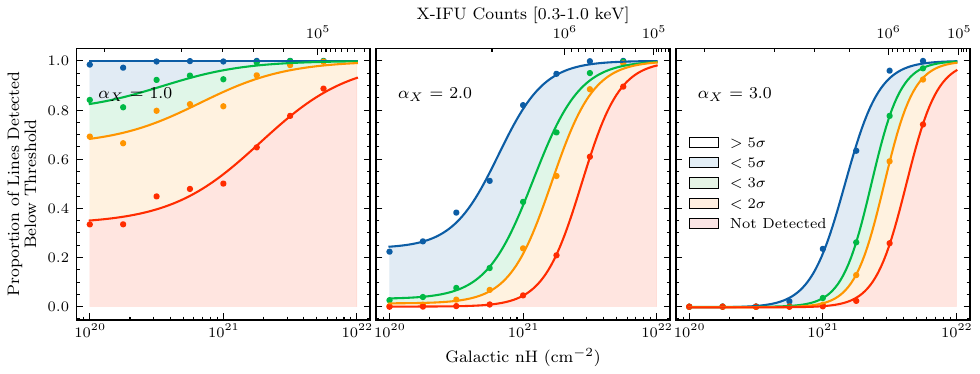}
        \caption{Proportion of lines that fall below 5$\sigma$, 3$\sigma$ and 2$\sigma$ thresholds, and lines that are not detected against local Galactic \nh{} and observed \gls*{xifu} counts in the 0.3--1.0\,keV range. The 2$\sigma$ threshold is included for comparison purposes. (Note: A threshold of $\sigma < 10^{-3}$ was used to determine a lack of detection)}
        \label{fig:line_proportions}
    \end{figure*}

    \subsection{Line significance measurements}\label{sec:line_sigs}

    The method to calculate the line significances in this investigation assumed that redshifts of \gls*{whim} absorbers are already known along the line of sight to the source. While this can be the case, it requires observations in different wavelengths (e.g. UV observations of \ion{O}{VI} absorption features or similar) to be completed before the \newathena{} observations. The upper portion of \Cref{fig:line_measures} displays the mean significance of detection, calculated using \Cref{eqn:deltaC}, across all redshifts as a function of \nh{} for the different \phoind{} values considered. These results were taken from the simulations outlined in \Cref{sec:first_sims}, with an assumed observation length of 50\,ks and unabsorbed source flux of ${F_{0.3-10.0\kev}}$\,=\,$5\times10^{-11}\funit$. Only lines with $z$\,>\,0.1 were considered for this analysis. A clear trend of decreasing measured line significance with increasing local \nh{} emerges for each \phoind, with lines at higher \nh{} values not being detected (seen in the gathering of distributions at $\sigma\sim0$). For \nh\,$\lesssim5\times10^{21}$\,cm$^{-2}$, an increase of source \phoind{} causes an increase of measured line significance. These trends suggest that \newathena{} should strive to observe sources with a lower local \nh{} along its line of sight when they attain a higher photon index to detect \gls*{whim} absorption features at a higher significance. This will also increase the number of counts detected by \newathena{}, reducing the required observation time of non-transient sources.
    
    Without prior detections to aid in predicting line positions, a blind line procedure must be carried out \citep[e.g.][]{Nicastro2018, Walsh2020}. While these methods are robust, statistical fluctuations must also be accounted for, and more importantly, ignored where applicable. To counter this, more ions expected to be present in the \gls*{whim} should be included in analysis, such as \ion{O}{VIII}, and other ionised species of C, N, Ne, and Fe. If two or more ions are detected at the same redshift, they can be considered as two tracers of the same filament, and thus their significances can be combined. However, this was not considered in this work. 

    \subsection{Equivalent width measurements}\label{sec:ew_recovery_results}

    When using absorption features to determine observable properties of \gls*{whim}, the line measurements must be accurate, for example, the \gls*{ew}. Using relative absorption feature strengths of different ions of the same atom, the temperature and density of the \gls*{whim} filament can be inferred, and hence accurate recovery of \glspl*{ew} is of utmost importance here. The lower portion of \Cref{fig:line_measures} displays how the rest-frame \glspl*{ew} were recovered for the ions with rest-frame \gls{ew} of 0.28\,eV, as outlined in \Cref{sec:first_sims}. For \phoind\,=\,1.0, the spread of measurements is the highest compared to the others, with spreads decreasing with increasing \phoind. The spread and hence the uncertainty in measurements also increases with increasing \nh, with very poor recovery for \nh\,$\gtrsim3\times10^{21}$\,cm$^{-2}$. Again, this reinforces that \newathena{} should aim to use sources with low \nh{} and high \phoind{} to increase the observed counts, and hence better detect and study \gls*{whim} features in absorption.
    
    \subsection{Line detection probabilities}

    Limits of observation to detect \gls*{whim} absorption features can also be derived by examining the proportions of absorption features that are detected at various significance levels as a function of \nh{} and counts detected in the 0.3--1.0\,keV band. \Cref{fig:line_proportions} displays this data at the $5\sigma$, $3\sigma$ and $2\sigma$ levels, as well as the proportion of lines that are not detected for each \phoind{} investigated, as outlined in \Cref{sec:first_sims}. These simulations assumed an observation length of 50\,ks and unabsorbed source flux of ${F_{0.3-10.0\kev}=5\times10^{-11}\funit}$. Only features with $z$\,>\,0.1 are studied here, as before in \Cref{sec:line_sigs,sec:ew_recovery_results}. An increase of proportions of lines that fall below these thresholds can be seen with an increase in Galactic \nh{} and a decrease in observed counts. The data were fit with truncated sigmoid functions to allow interpolation to calculate probabilistic limits. The ranges of the sigmoid functions were restricted to [$a$,1], where $a$ is found during the fitting process. This accounts for the probabilistic nature of line detection, allowing non-detections of lines in the low-\nh{} regime. We focus on the 50\%, 80\% and 90\% chance of detection limits at the 3$\sigma$ and 5$\sigma$ levels. These are only calculated for \phoind\,=\,2.0,\,3.0 as these limits are not reached for the \phoind\,=\,1.0 case. \Cref{tab:line_probabilities} displays the required millions of counts to detect an \ion{O}{VII} line at these probability limits. For the same probabilities of detection and the same significance of detection, there is very good agreement between the two \phoind{} values. It is clear that $\sim2.2\times10^6$ counts are required in the 0.3-1.0\,keV range for a 90\% chance of detecting an absorption feature of rest-frame \gls*{ew} of 0.28\,eV at a 3$\sigma$ level, while $\sim3.9\times10^6$ counts are required for a $5\sigma$ detection, however this value is not reached for \phoind\,=\,2.0. 

    To verify these calculations, we compare these thresholds to the previous count limits determined in \Cref{sec:requiredCounts}. To do this, the required counts for each level of significance were determined to have a one-tailed 1$\sigma$ probability of detecting the lines. For \phoind{} values of 2.0 and 3.0, these were calculated to be $9.00\times10^5$ and $8.41\times10^5$ for a 3$\sigma$ detection, and $1.52\times10^6$ and $1.34\times10^6$ for a 5$\sigma$ detection. From \Cref{fig:counts_for_sigs}, it can be seen that $\sim$\,7.41\,--\,8.92 $\times10^5$ and $\sim$\,1.40\,--\,1.66 $\times10^6$ counts are required for a detection at the same levels. The close agreement in these values indicates that the number of counts obtained in the $0.3-1.0\,\kev$ band is the prime proxy that determines the level of significance at which these lines are detected. In \Cref{sec:count_limits_first_results}, we provided the different count requirements for lines with varied rest-frame \glspl{ew} in the $z$\,=\,0.110\,--\,0.125 range. We did not complete a full suite of simulations of these lines outside this region, and hence we cannot carry out the same analysis as done here for lines with \gls*{ew}\,=\,0.28\,eV. However, given the good agreement of the values as previously discussed, the count requirements as displayed in \Cref{fig:varied_ew_counts} can be considered as accurate calculations as such.

    \begin{table}[t]
        \centering
        \caption{Millions of counts in the 0.3--1.0\,keV range required to obtain a 50\%, 80\% and 90\% probability of detection for an \ion{O}{VII} absorption feature at 3$\sigma$ and 5$\sigma$.}
        \label{tab:line_probabilities}
        \begin{tabular}{l||lll|lll}
        \hline
            \multirow{2}{*}{\phoind} & \multicolumn{3}{c|}{3$\sigma$} & \multicolumn{3}{c}{5$\sigma$}  \\ \cline{2-7} 
                 & 50\%     & 80\%     & 90\%     & 50\%     & 80\%     & 90\%     \\ \hline\hline
            2.0  & 1.48     & 1.95     & 2.20     & 2.09     & -        & - \\
            3.0  & 1.40     & 1.95     & 2.29     & 2.12     & 2.81     & 3.91 \\ \hline
        \end{tabular}
    \end{table}
    
    While we only study \ion{O}{VII} absorption features in this investigation, which fall between $\sim0.38-0.58\,\kev$, the counts observed in this range could serve as a better proxy. However, when considering all possible \gls*{whim} transitions, the $0.3-1.0\,\kev$ will include all features to be considered, and hence serve as a better proxy for all ions.

    \subsection{Calibration effects}

    Due to the ongoing development of the \gls*{xifu} instrument, the results presented in this investigation should be taken as a best-case scenario. As is the case with \xmm's \gls*{epic}, the spectrum becomes very poorly defined below $\sim0.5\,\kev$ due to detector noise. The variation in count rate of each energy band in the $0.5-12.0\,\kev$ band remains approximately constant at $\sim8\%$, while below $0.4\,\kev$, it reaches $\sim14\%$ \citep{Katayama2004}. While the current requirement for the knowledge of the energy scale is 0.5\,eV \citep[see][]{Peille2025}, due to contamination below $0.5\,\kev$ it is possible that the uncertainties in the absolute calibration may be as high as 10-20$\%$. Moreover, it is a possibility that more features from edges of materials not previously accounted for will be present below $0.5 \,\kev$. A combination of these effects will cause difficulties when detecting \gls*{whim} features in absorption within the key energy range of $<0.5\,\kev$.

    \subsection{Applicability for transient sources}\label{sec:grb_usage}

    While this work does not directly analyse transient sources, in particular \glspl{grb}, as a background X-ray beacon for detecting \gls*{whim} features in absorption, it does provide groundwork for the updated \newathena{} configuration for doing so. The main proxy for detecting these features, regardless of temporal nature, remains the same: the counts detected in the band of interest. In this work, we consider the soft X-ray band of energy range 0.3\,--\,1.0\,keV due to the number of X-ray transitions from ions present in the \gls{whim} falling in this band. As mentioned in \Cref{sec:introduction}, the new \gls{too} response time of \newathena{} has increased to 12 hours. While this increase will prove problematic for detecting absorption features in the spectra of \glspl{grb}, it may still be possible to do so with sufficiently bright sources. The count limits derived in this work, combined with early flux measurements of \glspl{grb}, can be used for determining if a transient event is bright enough to provide enough counts for a significant detection.

%%%%%%%%%%%%%%%%%%%%%%%%%%%%%%%%%%%%%%%%%%%%%%%%%%%%%%%%%%%%%%%%%%%%%%%%%%%%
% Conclusions
%%%%%%%%%%%%%%%%%%%%%%%%%%%%%%%%%%%%%%%%%%%%%%%%%%%%%%%%%%%%%%%%%%%%%%%%%%%%

\section{Conclusions}\label{sec:conclusions}

This work has investigated the capability of \newathena{} to accurately detect missing baryons residing in the \gls*{whim}, observed through single-line \ion{O}{VII}-He$\alpha$ absorption features in the continuum spectra of bright background power-law sources. The feature significances were calculated for these absorption features across a wide range of varied line redshift, local Galactic absorption, source photon index and observing length, with the assumption that the line positions were known a priori. The results of this work are summarised as follows:

   \begin{enumerate}
   
      \item Using a multidimensional approach, for a fixed unabsorbed source flux of $F_{0.3-10.0\,\text{keV}}=5\times10^{-11}\funit$ and observation length of 50\,ks, the redshift region corresponding to the highest usable cold Galactic \nh{} value for a 3$\sigma$ a priori single-line detection, \nhsig, was found to be $z$\,=\,0.110\,--\,0.125. This was done by imprinting a line of fixed rest-frame \gls*{ew} of 0.28\,eV, representing a mid-strength \gls*{whim} absorption feature, at different redshifts between 0.0 and 0.5, and varying source parameters, including cold Galactic absorption and photon index. For each set of parameters, 25 lines were simulated and their significances recorded. For each redshift region of size $dz$\,=\,0.005 and photon index, the measured significance was fit as a decaying exponential function of absorption, and hence the \nh{} corresponding to a 3$\sigma$ detection was found, as seen in \Cref{fig:zregion_vs_galnh_50ks}. This source flux and observation length allowed significant detections to be made for \phoind\,=\,2.0 and 3.0, but no significant detections were found for \phoind\,=\,1.0. Warmer phases of the \gls{igm} have not been included in this work, but known lines that can affect the detection of \gls{whim} absorption features have been included in the upper portion of \Cref{fig:zregion_vs_galnh_50ks}. No such lines fall within the redshift region of interest, and hence do not affect any further calculations in this work.
      
      \item For absorption features within the $z$\,=\,0.110\,--\,0.125 region, the number of counts required to obtain a 3$\sigma$ and a $5\sigma$ detection were found by varying the source \phoind{} between 1.0\,--\,3.0, the \nh{} along the line of sight to the source between $10^{20}$\,--\,$10^{22}$\,cm$^{-2}$, and the observation length between 6.25\,--\,150.0\,ks. It was found that $\sim$\,7.41\,--\,8.92 $\times10^5$ and $\sim$\,1.40\,--\,1.66 $\times10^6$ counts are required for a $3\sigma$  and 5$\sigma$ detection, as seen in \Cref{fig:counts_for_sigs}. This process was repeated for lines in the same redshift region with varied rest-frame \glspl{ew} between 0.14\,--\,0.42\,eV. The corresponding count requirements for these lines can be seen in \Cref{fig:varied_ew_counts}.

      \item A sample of known non-transient X-ray background sources was analysed here to assess their suitability for use in detecting and studying \gls*{whim} features in absorption. We utilised all publicly available data from the \swift archive to obtain the time-averaged spectrum for each source. Each spectrum was fit, and the two-dimensional relation represented by the 3$\sigma$ ellipse between flux and photon index was found using \verb|steppar|. Along this ellipse, the source states corresponding to the minimum and maximum \swift count rates were extracted. Ten 25\,ks observations were simulated for each of the best fit, minimum and maximum count parameters to obtain the expected \gls{xifu} count rates for each source. The merit for each source considered was then calculated, which can be seen in \Cref{fig:merit_sources}. The full list of sources considered can be seen in \Cref{tab:full_sources_table}.

      \item The measured single-line a priori significance averaged across all redshifts between 0.0 and 0.5 for an observation length of 50\,ks and unabsorbed source flux of $F_{0.3-10.0\,\text{keV}}=5\times10^{-11}\funit$ was found to decrease with increasing Galactic absorption, and increase with increasing source photon index, as seen in the top plot of \Cref{fig:line_measures}. For the case of \phoind\,=\,1.0, no significant lines were consistently detected, while a clear decreasing trend can be seen for \phoind\,=\,2.0 and 3.0. This reinforces that a higher photon index and lower Galactic absorption should be favoured when selecting a source to use as a background beacon to detect \gls*{whim} absorption features. 

      \item The recovery of the measured rest-frame \glspl*{ew} was investigated, as displayed in \Cref{fig:line_measures}. A higher \phoind{} and lower \nh{} are required to allow better measurements with minimal spread. This is clearly demonstrated for a \phoind\,=\,3.0 for lower absorption values. With decreasing \phoind, the spread of the measurements increases. For increasing values of \nh, the spread also increases, and for \nh\,$\gtrsim3\times10^{21}$\,cm$^{-2}$, the line width recovery becomes extremely poor. No sources above this \nh{} should be considered as a background X-ray source for studying \gls*{whim} features in absorption. 

      \item Across fixed values of \phoind{} and flux in the 0.3\,--\,10.0\,keV range, an increase in Galactic \nh{} causes a decrease in counts detected in the 0.3-1.0 keV band, and hence lowers the measured line significances. \Cref{fig:line_proportions} displays the proportions of lines that fall below $5\sigma$, $3\sigma$ and $2\sigma$ thresholds, as well as lines which are deemed undetected. Using these proportions, Galactic \nh{} limits for each \phoind, and hence count limits, were derived to obtain a 50\%, 80\% and 90\% probability of detecting lines at a $3\sigma$ and $5\sigma$ detection across all redshifts that are not affected by absorption and instrumental features. These count limits are shown in \Cref{tab:line_probabilities}. It was found that these limits agree with those previously used to calculate the required observing lengths. 
      
   \end{enumerate}

\begin{acknowledgements}
    J.F. and A.M.C. acknowledge support from the European Space Agency (PRODEX) (Grant No. 4000138314). J.S. acknowledges support from Netherlands Organization for Scientific Research (NWO) through research programme Athena 184.034.002. We acknowledge the use of public data from the Swift data archive. This research has made use of data and/or software provided by the High Energy Astrophysics Science Archive Research Center (HEASARC), which is a service of the Astrophysics Science Division at NASA/GSFC. We would like to thank the referee for their constructive input and guidance in this work.
\end{acknowledgements}

\bibliographystyle{aa}
\bibliography{athena.bib}

\begin{appendix}
    \nolinenumbers
    \onecolumn
    
    \section{Galactic absorption corresponding to $3\sigma$ detections}\label{sec:galnh_appendix}
    
    \begin{figure*}[h]
        \centering
        \includegraphics{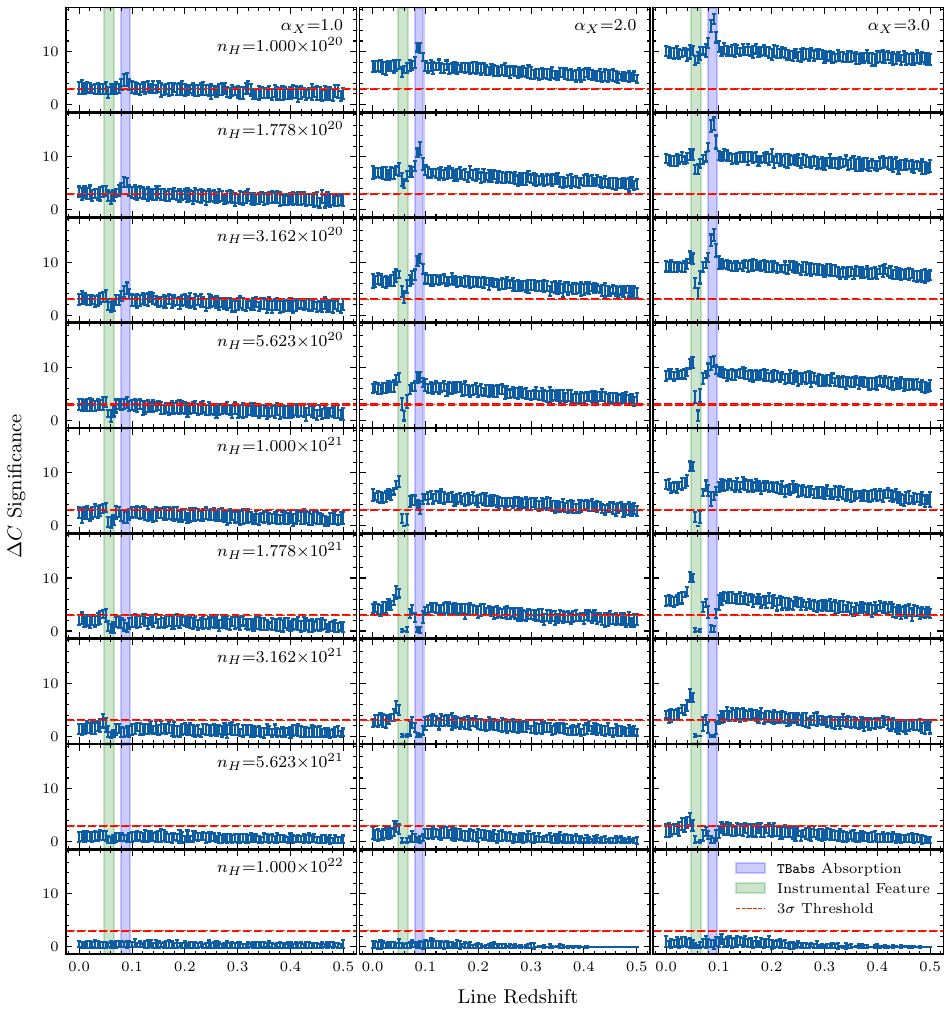}
        \caption{Measured line significances for all simulation parameters. From left to right, the columns display the data for \phoind{} of 1.0, 2.0 and 3.0, respectively. From top to bottom, the rows display the data for \nh{} between $10^{20}$\,--\,$10^{22}$\,cm$^{-2}$. As before the local absorption and instrumental features are highlighted by blue and green regions, respectively. The 3$\sigma$ threshold is denoted by a horizontal red dotted line.}
        \label{fig:full_sigs_results}
    \end{figure*}
    
        Here, we display the full data used to calculate the Galactic absorption values that correspond to a 3$\sigma$ detection, \nhsig. These values have been calculated assuming an unabsorbed source flux of $F_{0.3-10.0\,\text{keV}}\,=\,5\times10^{-11}\funit$ and a 50\,ks observation length. \Cref{fig:full_sigs_results} displays the measured line significances for all simulation parameters, where each data point represents the 3$\sigma$-clipped Gaussian spread of 25 runs. 
    
        The data displayed in \Cref{fig:zregion_vs_galnh_50ks} were obtained by fitting decaying exponential functions to measured line significances of the same \phoind{} and $z$ against \nh{} using Monte Carlo methods. From these fits, \nhsig{} could be extracted. Through inspection of \Cref{fig:full_sigs_results}, the effect of the local absorption and instrumental features can be easily understood: For all \phoind{} and \nh, the instrumental feature (green region) causes a reduction in the measured significance. However, the local absorption feature (blue region) causes an increase of significance for \nh\,$\lesssim10^{21}\cm^{-2}$. However, for higher absorptions, the effect reverses and causes a reduction in measured significance. For \phoind{} values 2.0 and 3.0, this causes a faster decaying fit, and hence the lower \nhsig{} values. Moreover, the apparent spike of \nhsig{} at redshifts on the lower boundary of the instrumental feature ($z \approx 0.05$) is a result of the high measured line significance, in particular for \nh$\gtrsim1\times10^{21}\cm^{-2}$, as seen in \Cref{fig:full_sigs_results}. For \phoind{} of 1.0, due to the lower measured significances for low absorptions ($\sim$3$\sigma$ for \nh\,=\,$1\times10^{20}\cm^{-2}$), the positive effect from the local absorption feature causes these features to be measured above the 3$\sigma$ threshold, as can be seen in \Cref{fig:zregion_vs_galnh_50ks}. It is clear that these $z$ ranges cause inaccurate measurements of the line significances, and hence these regions should be avoided where possible when detecting \gls*{whim} features in absorption. 
    
    \section{Sources analysed in this work}\label{sec:full_sources_appendix}
    
    \Cref{fig:appendix_merit_sources} displays the full version of \Cref{fig:merit_sources}, ordered by source merit. \Cref{tab:full_sources_table} summarises all the observable sources considered in this work, ordered by source position. These sources were chosen from the sources outlined in \Cref{sec:real_sources}, provided they had a measured redshift of $z$\,>\,0.1. The first three columns of \Cref{tab:full_sources_table} provide the source name and position; Columns 4-5 provide the cold local Galactic absorption along the line of sight to the source provided by the HEASARC \nh{} Column Density Tool and the measured redshift of the source; Columns 6-7 display the fitted source photon index and flux in the 0.3\,--\,10.0\,keV range, in units of $1\times 10^{-12} \funit$, extracted from the \swift data; Column 8 provides the calculated merit of each source, displayed in \Cref{fig:merit_sources}; and column 9 provides the references from which the source was taken.
    
        \begin{figure*}[h]
            \centering
            \includegraphics[width=\linewidth]{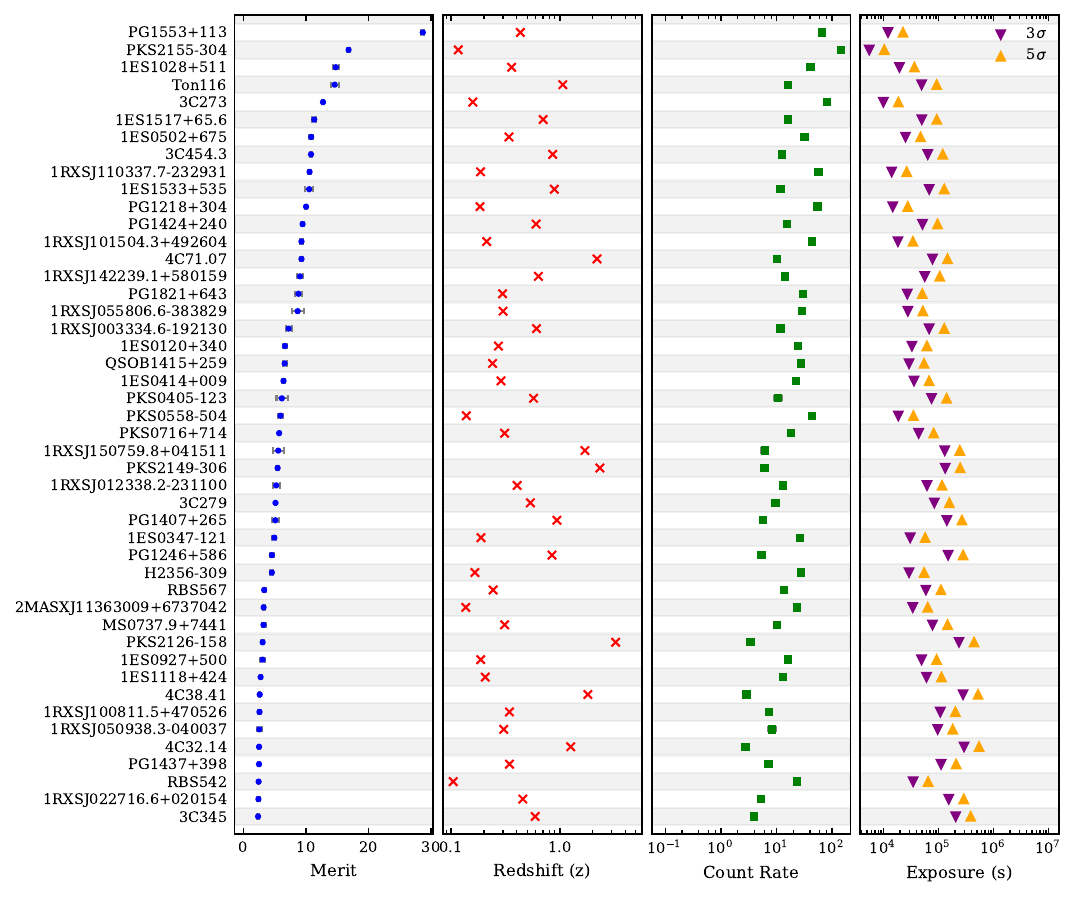}
            \caption{Merit, source redshift and measured \newathena{} count rate of each source investigated in this work. The rightmost portion displays the required observation length to detect a feature in absorption, using the count limits from \Cref{fig:counts_for_sigs}. The purple downward triangles and orange upward triangles display the required exposure for a 3$\sigma$ and 5$\sigma$ detection of a line of rest-frame \gls{ew} of 0.28\,eV falling in the $z$\,=\,0.110\,--\,0.125 region, respectively. All data can be seen in \Cref{tab:full_sources_table}.}
            \label{fig:appendix_merit_sources}
        \end{figure*}
        \FloatBarrier
        \begin{figure*}[!t]\ContinuedFloat
            \centering
            \includegraphics[width=\linewidth]{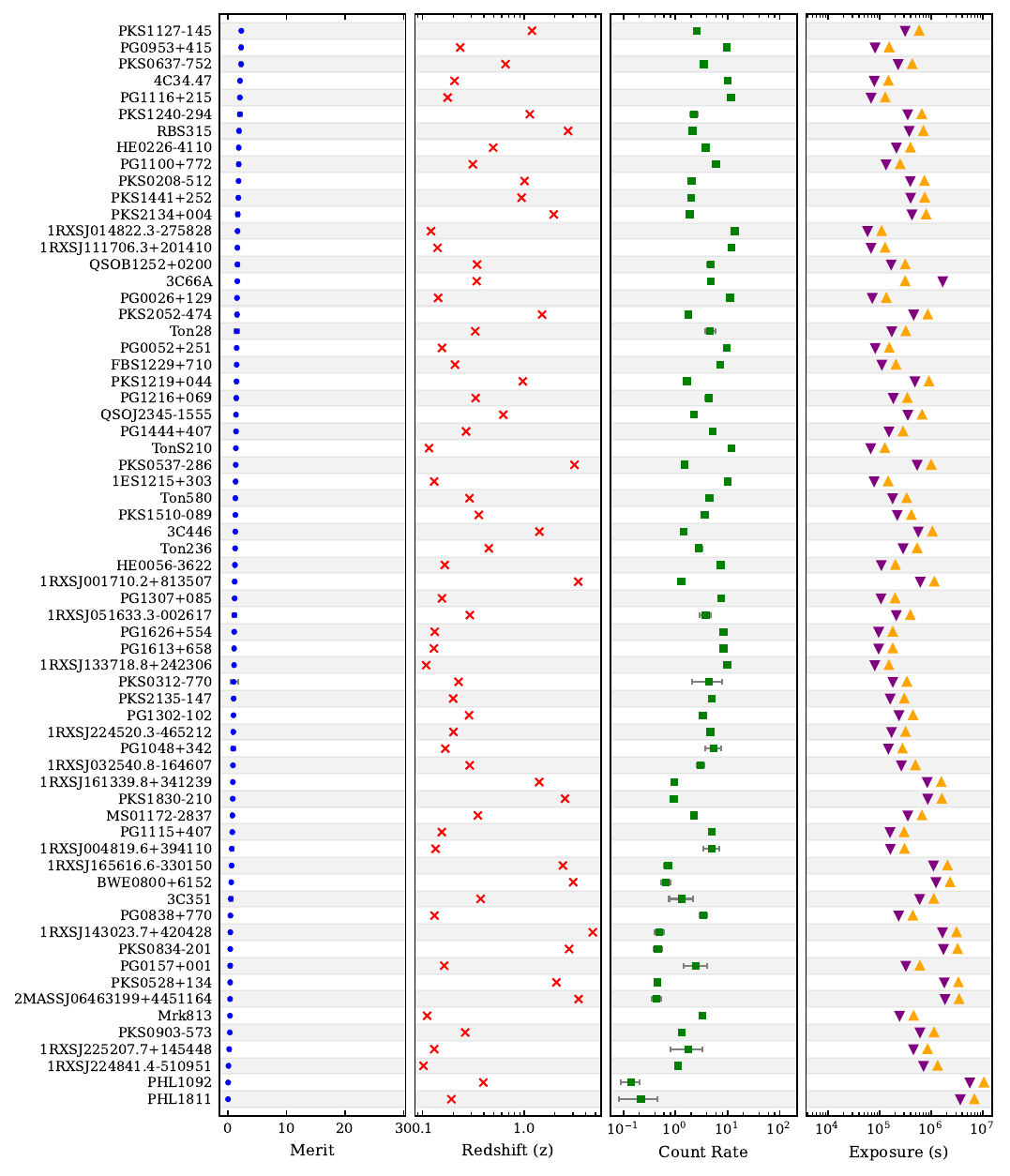}
            \caption{continued.}
        \end{figure*}
    
\longtab[2]{
\small{
\begin{longtable}{l||c|c|c|c|c|c|c|c}

\caption{All non-transient power-law sources considered in this work}\label{tab:full_sources_table}\\
\hline
Source & RA & Dec & $z$ & \phoind{} & $F_X$ & \nh{} & Merit & Refs. \\ \hline \hline
\endfirsthead
\caption{continued.}\\
\hline
Source & RA & Dec & $z$ & \phoind{} & $F_X$ & \nh{} & Merit & Refs. \\ \hline \hline
\endhead
1RXSJ001710.2+813507 & 00 17 08.5 & +81 35 08.1 & 3.378 & 1.23$\pm$0.05 & 6.4$\pm$0.3 & 1.29E+21 & 1.17 & 7 \\
PG0026+129 & 00 29 13.7 & +13 16 04.0 & 0.142 & 1.99$\pm$0.03 & 13.5$\pm$0.3 & 4.68E+20 & 1.57 & 3 \\
1RXSJ003334.6-192130 & 00 33 34.4 & -19 21 33.1 & 0.610 & 2.24$\pm$0.04 & 9.4$\pm$0.3 & 1.52E+20 & 7.21 & 6 \\
1RXSJ004819.6+394110 & 00 48 19.0 & +39 41 11.7 & 0.134 & 1.9$\pm$0.2 & 6.4$\pm$1.0 & 4.67E+20 & 0.66 & 9 \\
PG0052+251 & 00 54 52.1 & +25 25 39.0 & 0.155 & 1.91$\pm$0.04 & 12.5$\pm$0.4 & 4.09E+20 & 1.5 & 3 \\
HE0056-3622 & 00 58 37.4 & -36 06 04.9 & 0.165 & 2.09$\pm$0.04 & 6.7$\pm$0.2 & 1.35E+20 & 1.22 & 9 \\
MS01172-2837 & 01 19 35.7 & -28 21 31.4 & 0.349 & 2.69$\pm$0.07 & 1.36$\pm$0.05 & 1.64E+20 & 0.79 & 6 \\
TonS210 & 01 21 51.5 & -28 20 57.7 & 0.116 & 2.34$\pm$0.07 & 8.8$\pm$0.4 & 1.62E+20 & 1.37 & 3,4,9 \\
1ES0120+340 & 01 23 08.6 & +34 20 48.7 & 0.272 & 1.79$\pm$0.02 & 37.4$\pm$0.6 & 4.63E+20 & 6.61 & 6,7 \\
1RXSJ012338.2-231100 & 01 23 38.3 & -23 10 58.9 & 0.404 & 1.92$\pm$0.06 & 13.9$\pm$0.7 & 1.19E+20 & 5.23 & 6 \\
PHL1092 & 01 39 55.8 & +06 19 22.5 & 0.395 & 5.0$\pm$0.8 & 0.08$\pm$0.02 & 3.29E+20 & 0.06 & 6 \\
1RXSJ014822.3-275828 & 01 48 22.3 & -27 58 25.6 & 0.121 & 2.91$\pm$0.02 & 7.45$\pm$0.10 & 1.47E+20 & 1.65 & 6 \\
PG0157+001 & 01 59 50.2 & +00 23 40.7 & 0.163 & 1.9$\pm$0.3 & 2.7$\pm$0.7 & 2.3E+20 & 0.4 & 9 \\
PKS0208-512 & 02 10 46.2 & -51 01 01.9 & 1.003 & 1.49$\pm$0.02 & 3.89$\pm$0.07 & 1.6E+20 & 1.82 & 7 \\
3C66A & 02 22 39.6 & +43 02 07.8 & 0.340\footnote{Updated redshift from \citet{TorresZafra2018, Jones2021}.} & 2.33$\pm$0.02 & 5.30$\pm$0.06 & 8.17E+20 & 1.76 & 8,9 \\
RBS315 & 02 25 04.7 & +18 46 48.8 & 2.690 & 1.05$\pm$0.04 & 12.3$\pm$0.4 & 9.5E+20 & 1.91 & 6,7 \\
1RXSJ022716.6+020154 & 02 27 16.6 & +02 02 00.0 & 0.456 & 2.00$\pm$0.06 & 5.6$\pm$0.2 & 2.74E+20 & 2.38 & 6 \\
HE0226-4110 & 02 28 15.2 & -40 57 14.8 & 0.494 & 2.37$\pm$0.05 & 2.70$\pm$0.08 & 1.45E+20 & 1.86 & 4,5,6,9 \\
PKS0312-770 & 03 11 55.2 & -76 51 50.8 & 0.226 & 1.9$\pm$0.4 & 7$\pm$2 & 7.86E+20 & 1 & 3,4,9 \\
1RXSJ032540.8-164607 & 03 25 41.1 & -16 46 16.9 & 0.290 & 3.0$\pm$0.1 & 1.8$\pm$0.1 & 3.29E+20 & 0.88 & 6 \\
4C32.14 & 03 36 30.1 & +32 18 29.3 & 1.259 & 1.12$\pm$0.03 & 15.7$\pm$0.4 & 1.22E+21 & 2.47 & 7 \\
1ES0347-121 & 03 49 23.2 & -11 59 27.4 & 0.188 & 1.88$\pm$0.03 & 33.1$\pm$0.9 & 2.98E+20 & 4.91 & 6,7 \\
PKS0405-123 & 04 07 48.4 & -12 11 36.7 & 0.573 & 1.81$\pm$0.08 & 15.0$\pm$1.0 & 3.66E+20 & 6.12 & 2,3,4,5,6 \\
1ES0414+009 & 04 16 52.5 & +01 05 23.9 & 0.287 & 2.22$\pm$0.03 & 26.4$\pm$0.5 & 7.72E+20 & 6.39 & 1,6 \\
RBS542 & 04 26 00.7 & -57 12 01.8 & 0.104 & 1.92$\pm$0.02 & 24.7$\pm$0.4 & 1.12E+20 & 2.41 & 6,9 \\
RBS567 & 04 39 38.7 & -53 11 31.2 & 0.243 & 3.12$\pm$0.05 & 6.5$\pm$0.2 & 6.22E+19 & 3.3 & 9 \\
1ES0502+675 & 05 07 56.2 & +67 37 24.2 & 0.340 & 1.93$\pm$0.02 & 52.8$\pm$0.6 & 9.15E+20 & 10.81 & 6 \\
1RXSJ050938.3-040037 & 05 09 38.2 & -04 00 45.7 & 0.304 & 1.76$\pm$0.09 & 15$\pm$1 & 7.92E+20 & 2.52 & 6 \\
1RXSJ051633.3-002617 & 05 16 33.3 & -00 26 17.0 & 0.291 & 2.0$\pm$0.1 & 5.8$\pm$0.6 & 9.92E+20 & 1.1 & 6 \\
PKS0528+134 & 05 30 56.4 & +13 31 55.1 & 2.064 & 1.11$\pm$0.03 & 3.91$\pm$0.09 & 2.36E+21 & 0.4 & 7 \\
PKS0537-286 & 05 39 54.3 & -28 39 55.9 & 3.106 & 1.16$\pm$0.03 & 5.0$\pm$0.1 & 1.81E+20 & 1.34 & 7 \\
1RXSJ055806.6-383829 & 05 58 06.4 & -38 38 31.6 & 0.300 & 2.11$\pm$0.06 & 29$\pm$1 & 3.05E+20 & 8.65 & 6 \\
PKS0558-504 & 05 59 47.4 & -50 26 52.0 & 0.138 & 2.29$\pm$0.04 & 37.3$\pm$1.0 & 3.28E+20 & 5.9 & 3,6,9 \\
PKS0637-752 & 06 35 46.5 & -75 16 16.8 & 0.653 & 1.56$\pm$0.03 & 8.1$\pm$0.2 & 7.21E+20 & 2.27 & 6,7 \\
2MASSJ06463199+4451164 & 06 46 32.0 & +44 51 16.6 & 3.408 & 1.5$\pm$0.1 & 1.3$\pm$0.1 & 1.07E+21 & 0.39 & 7 \\
PKS0716+714 & 07 21 53.4 & +71 20 36.4 & 0.310 & 2.187$\pm$0.008 & 16.80$\pm$0.09 & 2.89E+20 & 5.69 & 6,7,8,9 \\
MS0737.9+7441 & 07 44 05.4 & +74 33 58.3 & 0.310 & 2.26$\pm$0.06 & 9.1$\pm$0.4 & 3.28E+20 & 3.19 & 6 \\
BWE0800+6152 & 08 05 18.3 & +61 44 26.2 & 3.018 & 1.08$\pm$0.09 & 2.8$\pm$0.3 & 4.39E+20 & 0.58 & 7 \\
PG0804+761 & 08 10 58.7 & +76 02 42.5 & 0.099 & 2.15$\pm$0.02 & 23.5$\pm$0.4 & 3.35E+20 & 2.39 & 4,6,9 \\
PKS0834-201 & 08 36 39.2 & -20 16 59.5 & 2.750 & 1.02$\pm$0.08 & 2.3$\pm$0.2 & 5.86E+20 & 0.41 & 7 \\
4C71.07 & 08 41 24.4 & +70 53 42.2 & 2.200 & 1.204$\pm$0.010 & 33.7$\pm$0.3 & 2.85E+20 & 9.26 & 6,7 \\
PG0838+770 & 08 44 45.3 & +76 53 09.5 & 0.131 & 2.14$\pm$0.10 & 3.1$\pm$0.2 & 2.34E+20 & 0.45 & 9 \\
PKS0903-573 & 09 04 53.2 & -57 35 05.8 & 0.262\footnote{Updated redshift from \citet{Goldoni2024}.} & 1.56$\pm$0.03 & 6.6$\pm$0.1 & 2.59E+21 & 0.35 & 7 \\
1ES0927+500 & 09 30 37.6 & +49 50 25.5 & 0.187 & 2.00$\pm$0.07 & 16.2$\pm$0.9 & 1.49E+20 & 3.03 & 6 \\
PG0953+415 & 09 56 52.4 & +41 15 22.2 & 0.234 & 2.20$\pm$0.03 & 7.8$\pm$0.2 & 1.09E+20 & 2.28 & 2,4,6,9 \\
Ton28 & 10 04 02.6 & +28 55 35.3 & 0.329 & 2.8$\pm$0.2 & 2.6$\pm$0.3 & 1.79E+20 & 1.52 & 4,6,9 \\
1RXSJ100811.5+470526 & 10 08 11.4 & +47 05 21.5 & 0.343 & 1.96$\pm$0.04 & 7.4$\pm$0.2 & 8.45E+19 & 2.54 & 6 \\
1RXSJ101504.3+492604 & 10 15 04.1 & +49 26 00.7 & 0.212 & 2.27$\pm$0.01 & 33.0$\pm$0.3 & 9.35E+19 & 9.26 & 6 \\
1ES1028+511 & 10 31 18.5 & +50 53 35.8 & 0.360 & 2.06$\pm$0.02 & 37.8$\pm$0.6 & 1.11E+20 & 14.76 & 1,6,7,9 \\
PG1048+342 & 10 51 43.9 & +33 59 26.7 & 0.167 & 2.2$\pm$0.2 & 4.6$\pm$0.8 & 1.74E+20 & 0.9 & 9 \\
1RXSJ110337.7-232931 & 11 03 37.6 & -23 29 31.2 & 0.186 & 1.91$\pm$0.01 & 78.5$\pm$0.7 & 5.13E+20 & 10.55 & 6,9 \\
PG1100+772 & 11 04 13.9 & +76 58 58.2 & 0.312 & 1.90$\pm$0.06 & 7.2$\pm$0.4 & 2.79E+20 & 1.86 & 4 \\
1RXSJ111706.3+201410 & 11 17 06.3 & +20 14 07.4 & 0.140 & 2.27$\pm$0.05 & 8.9$\pm$0.3 & 1.25E+20 & 1.64 & 6 \\
PG1115+407 & 11 18 30.3 & +40 25 54.1 & 0.154 & 2.29$\pm$0.04 & 3.8$\pm$0.1 & 1.47E+20 & 0.77 & 9 \\
PG1116+215 & 11 19 08.7 & +21 19 18.0 & 0.176 & 2.20$\pm$0.04 & 9.4$\pm$0.3 & 1.19E+20 & 2.05 & 2,3,4,5,6,9 \\
1ES1118+424 & 11 20 48.1 & +42 12 12.5 & 0.206\footnote{Lower limit from \citet{Jones2021}.} & 2.37$\pm$0.03 & 9.7$\pm$0.2 & 1.74E+20 & 7.51 & 8 \\
PKS1127-145 & 11 30 07.1 & -14 49 27.4 & 1.188 & 1.13$\pm$0.02 & 9.7$\pm$0.2 & 3.47E+20 & 2.3 & 7 \\
Ton580 & 11 31 09.5 & +31 14 05.5 & 0.289 & 1.95$\pm$0.05 & 4.8$\pm$0.2 & 1.88E+20 & 1.29 & 9 \\
2MASXJ11363009+6737042 & 11 36 30.1 & +67 37 04.3 & 0.136 & 1.71$\pm$0.02 & 33.5$\pm$0.7 & 1.43E+20 & 3.2 & 6 \\
1ES1215+303 & 12 17 52.1 & +30 07 00.6 & 0.130 & 2.48$\pm$0.03 & 7.0$\pm$0.1 & 1.82E+20 & 1.32 & 6,8 \\
PG1216+069 & 12 19 20.9 & +06 38 38.5 & 0.332 & 2.05$\pm$0.09 & 4.1$\pm$0.3 & 1.57E+20 & 1.44 & 3,4,9 \\
PG1218+304 & 12 21 21.9 & +30 10 37.2 & 0.184 & 1.91$\pm$0.01 & 62.7$\pm$0.6 & 1.91E+20 & 9.99 & 6,7 \\
PKS1219+044 & 12 22 22.6 & +04 13 13.0 & 0.966 & 1.26$\pm$0.05 & 4.5$\pm$0.2 & 1.58E+20 & 1.48 & 7 \\
3C273 & 12 29 06.7 & +02 03 08.6 & 0.158 & 1.533$\pm$0.006 & 151.4$\pm$0.8 & 1.7E+20 & 12.71 & 3,4,5,6,8,9 \\
FBS1229+710 & 12 31 36.6 & +70 44 14.4 & 0.208 & 1.76$\pm$0.03 & 9.5$\pm$0.2 & 1.56E+20 & 1.5 & 6 \\
PKS1240-294 & 12 43 10.7 & -29 43 22.5 & 1.130 & 1.41$\pm$0.08 & 6.1$\pm$0.4 & 5.52E+20 & 2.05 & 7 \\
Ton116 & 12 43 12.7 & +36 27 44.0 & 1.065 & 2.30$\pm$0.03 & 12.2$\pm$0.2 & 1.32E+20 & 14.57 & 6 \\
PG1246+586 & 12 48 18.8 & +58 20 28.7 & 0.847 & 2.41$\pm$0.04 & 3.6$\pm$0.1 & 7.65E+19 & 4.53 & 6 \\
QSOB1252+0200 & 12 55 19.7 & +01 44 12.2 & 0.343 & 2.02$\pm$0.09 & 4.6$\pm$0.3 & 1.36E+20 & 1.63 & 6 \\
3C279 & 12 56 11.2 & -05 47 21.5 & 0.535 & 1.513$\pm$0.006 & 18.5$\pm$0.1 & 2.25E+20 & 5.11 & 1,6,7 \\
PG1302-102 & 13 05 33.0 & -10 33 19.4 & 0.286 & 1.83$\pm$0.02 & 4.48$\pm$0.06 & 3.09E+20 & 0.96 & 3,4,9 \\
PG1307+085 & 13 09 47.0 & +08 19 48.2 & 0.155 & 2.11$\pm$0.04 & 7.0$\pm$0.2 & 2.07E+20 & 1.16 & 3,9 \\
1RXSJ133718.8+242306 & 13 37 18.7 & +24 23 03.3 & 0.109 & 2.02$\pm$0.02 & 9.5$\pm$0.2 & 1.1E+20 & 1.07 & 6 \\
PG1407+265 & 14 09 23.9 & +26 18 21.1 & 0.940 & 2.17$\pm$0.07 & 4.7$\pm$0.2 & 1.15E+20 & 5.07 & 1,6 \\
QSOB1415+259 & 14 17 56.7 & +25 43 26.2 & 0.240 & 1.92$\pm$0.03 & 30.5$\pm$0.7 & 1.68E+20 & 6.58 & 6 \\
1RXSJ142239.1+580159 & 14 22 38.9 & +58 01 55.5 & 0.635 & 1.94$\pm$0.03 & 14.6$\pm$0.3 & 9.35E+19 & 9.02 & 6 \\
PG1424+240 & 14 27 00.4 & +23 48 00.0 & 0.605 & 2.44$\pm$0.02 & 12.2$\pm$0.1 & 3.39E+20 & 9.44 & 8 \\
Mrk813 & 14 27 25.0 & +19 49 52.3 & 0.111 & 1.47$\pm$0.07 & 6.7$\pm$0.4 & 2.26E+20 & 0.36 & 6 \\
1RXSJ143023.7+420428 & 14 30 23.7 & +42 04 36.5 & 4.700 & 1.23$\pm$0.09 & 1.3$\pm$0.1 & 9.28E+19 & 0.43 & 7 \\
PG1437+398 & 14 39 17.5 & +39 32 42.8 & 0.344 & 2.54$\pm$0.05 & 4.5$\pm$0.1 & 1.04E+20 & 2.47 & 6 \\
PKS1441+252 & 14 43 56.9 & +25 01 44.5 & 0.940 & 2.05$\pm$0.05 & 2.07$\pm$0.08 & 3.06E+20 & 1.8 & 7 \\
PG1444+407 & 14 46 45.9 & +40 35 05.8 & 0.267 & 2.49$\pm$0.08 & 3.4$\pm$0.2 & 1.09E+20 & 1.4 & 4,9 \\
1RXSJ150759.8+041511 & 15 07 59.7 & +04 15 12.0 & 1.703 & 2.4$\pm$0.1 & 4.9$\pm$0.3 & 3.6E+20 & 5.54 & 6 \\
PKS1510-089 & 15 12 50.5 & -09 05 59.8 & 0.356 & 1.315$\pm$0.008 & 12.14$\pm$0.09 & 7.16E+20 & 1.29 & 7 \\
1ES1517+65.6 & 15 17 47.6 & +65 25 23.3 & 0.702 & 1.99$\pm$0.02 & 17.4$\pm$0.2 & 2.58E+20 & 11.3 & 6 \\
Ton236 & 15 28 40.6 & +28 25 29.8 & 0.447 & 1.98$\pm$0.08 & 2.9$\pm$0.2 & 2.15E+20 & 1.25 & 5 \\
1ES1533+535 & 15 35 00.8 & +53 20 37.3 & 0.890 & 2.10$\pm$0.04 & 10.6$\pm$0.3 & 1.39E+20 & 10.5 & 6 \\
PG1553+113 & 15 55 43.0 & +11 11 24.4 & 0.433 & 2.197$\pm$0.005 & 63.8$\pm$0.2 & 3.61E+20 & 23.83 & 3,6,8,9 \\
1RXSJ161339.8+341239 & 16 13 41.1 & +34 12 47.9 & 1.398 & 1.57$\pm$0.07 & 1.6$\pm$0.1 & 1.42E+20 & 0.86 & 7 \\
PG1613+658 & 16 13 57.2 & +65 43 10.0 & 0.129 & 1.875$\pm$0.009 & 10.05$\pm$0.07 & 2.37E+20 & 1.07 & 5,6,9 \\
PG1626+554 & 16 27 56.1 & +55 22 31.6 & 0.132 & 2.16$\pm$0.03 & 6.9$\pm$0.2 & 1.07E+20 & 1.09 & 9 \\
4C38.41 & 16 35 15.5 & +38 08 04.5 & 1.814 & 1.43$\pm$0.02 & 5.82$\pm$0.09 & 9.7E+19 & 2.58 & 7 \\
3C345 & 16 42 58.8 & +39 48 37.0 & 0.593 & 1.60$\pm$0.02 & 6.2$\pm$0.1 & 8.86E+19 & 2.32 & 7 \\
1RXSJ165616.6-330150 & 16 56 16.8 & -33 02 12.7 & 2.392 & 0.95$\pm$0.07 & 8.4$\pm$0.5 & 2.35E+21 & 0.65 & 7 \\
3C351 & 17 04 41.4 & +60 44 30.5 & 0.372 & 1.4$\pm$0.3 & 3.0$\pm$0.7 & 1.86E+20 & 0.49 & 4 \\
4C34.47 & 17 23 20.8 & +34 17 58.0 & 0.206 & 1.74$\pm$0.03 & 14.8$\pm$0.4 & 2.97E+20 & 2.08 & 6,7 \\
PG1821+643 & 18 21 57.2 & +64 20 36.2 & 0.297 & 1.66$\pm$0.03 & 50$\pm$1 & 3.46E+20 & 8.78 & 2,4,5,6,9 \\
PKS1830-210 & 18 33 39.9 & -21 03 39.4 & 2.507 & 0.54$\pm$0.02 & 19.5$\pm$0.3 & 1.88E+21 & 0.84 & 6,7 \\
PKS2052-474 & 20 56 16.4 & -47 14 47.6 & 1.493 & 1.35$\pm$0.05 & 4.4$\pm$0.2 & 2.7E+20 & 1.57 & 7 \\
PKS2126-158 & 21 29 12.2 & -15 38 41.0 & 3.268 & 1.21$\pm$0.03 & 11.6$\pm$0.3 & 4.09E+20 & 3.06 & 6,7 \\
PKS2134+004 & 21 36 38.6 & +00 41 54.2 & 1.941 & 1.53$\pm$0.06 & 4.0$\pm$0.2 & 4.47E+20 & 1.69 & 7 \\
PKS2135-147 & 21 37 45.2 & -14 32 55.7 & 0.200 & 1.50$\pm$0.05 & 10.8$\pm$0.5 & 4.17E+20 & 0.99 & 1,6 \\
PKS2149-306 & 21 51 55.5 & -30 27 53.7 & 2.340 & 1.15$\pm$0.02 & 20.2$\pm$0.4 & 1.62E+20 & 5.44 & 7 \\
PHL1811 & 21 55 01.5 & -09 22 24.3 & 0.192 & 1.4$\pm$0.4 & 0.5$\pm$0.2 & 4.01E+20 & 0.04 & 3,4,5,9 \\
PKS2155-304 & 21 58 52.1 & -30 13 32.1 & 0.116 & 2.411$\pm$0.007 & 104.3$\pm$0.5 & 1.28E+20 & 16.79 & 3,4,5,6,8,9 \\
3C446 & 22 25 47.3 & -04 57 01.4 & 1.404 & 1.50$\pm$0.06 & 3.2$\pm$0.2 & 4.97E+20 & 1.28 & 7 \\
1RXSJ224520.3-465212 & 22 45 20.3 & -46 52 11.6 & 0.201 & 2.30$\pm$0.06 & 3.4$\pm$0.1 & 8.23E+19 & 0.94 & 6,9 \\
1RXSJ224841.4-510951 & 22 48 41.2 & -51 09 53.3 & 0.102 & 1.16$\pm$0.03 & 3.5$\pm$0.1 & 8.64E+19 & 0.11 & 6,9 \\
1RXSJ225207.7+145448 & 22 52 07.7 & +14 54 48.0 & 0.130 & 2.0$\pm$0.4 & 2.1$\pm$0.7 & 4.6E+20 & 0.23 & 6 \\
3C454.3 & 22 53 57.7 & +16 08 53.6 & 0.859 & 1.339$\pm$0.007 & 40.1$\pm$0.3 & 6.76E+20 & 10.79 & 6,7 \\
QSOJ2345-1555 & 23 45 12.5 & -15 55 07.8 & 0.621 & 1.95$\pm$0.04 & 2.40$\pm$0.07 & 1.71E+20 & 1.4 & 9 \\
H2356-309 & 23 59 07.9 & -30 37 40.7 & 0.165 & 1.90$\pm$0.03 & 30.3$\pm$0.7 & 1.33E+20 & 4.52 & 6,8,9
\end{longtable}
\tablebib{(1) \cite{Fang2005}; (2) \cite{Danforth2010}; (3) \cite{Prochaska2011}; (4) \cite{Tilton2012}; (5) \cite{Savage2014,Stocke2014}; (6) \cite{Bregman2015}; (7) \cite{Dalton2021}; (8) \cite{Jones2021}; (9) \cite{Spence2025}}
}
}

\end{appendix}

\end{document}